# Universality of the 1/9 Magnetization Plateau and Quantum-Disordered States in the Kagome Family $Cs_8AB_3Ti_{12}F_{48}$ ($A$ = Rb, Li; $B$ = K, Na)

Prena Chaudhary[1,13], Asiri Thennakoon[1,13], Tommy Park[1], Hanru Wang[2], Leshan Zhao[3], Laurel Winter[4], Neil Herrison[4], Christina Hoffmann[5], Junghong H. He[5], Harald O. Jeschke[6], Hiroyuki Nojiri[7], Akira Matsuo[8], Koichi Kindo[8], Miwako Takahashi[9], Yukio Noda[8,10], Taku J. Sato[8,10,11], Shiyan Li[2], Hiroaki Ueda[12]✉, Gia-Wei Chern[1], and Seung-Hun Lee[1]✉

[1]Department of Physics, University of Virginia; Charlottesville, Virginia, 22904, USA.

[2]State Key Laboratory of Surface Physics, Department of Physics, Fudan University, Shanghai 200438, China

[3]Institute for Quantum Matter and Department of Physics and Astronomy, The Johns Hopkins University; Baltimore, Maryland, 21218, USA.

[4]National High Magnetic Field Laboratory, Los Alamos National Laboratory, Los Alamos, NM, 87545, USA

[5]Oak Ridge National Laboratory, Oak Ridge, Tennessee, 37831, USA.

[6]Research Institute for Interdisciplinary Science, Okayama University, Okayama 700-8530, Japan

[7]Institute for Materials Research, Tohoku University, Sendai 980-8577, Japan

[8]Institute for Solid State Physics, University of Tokyo, Kashiwa, 277-8581, Japan

[9]Department of Materials Science, Institute of Pure and Applied Sciences, University of Tsukuba, Tsukuba, Ibaraki, 305-8573, Japan

[10]Institute of Multidisciplinary Research for Advanced Materials, Tohoku University, Sendai 980-8577, Japan

[11]Trans-scale Quantum Science Institute, University of Tokyo, Tokyo 113-0033, Japan

[12]Co-Creation Institute for Advanced Materials, Shimane University, 1060 Nishikawatsu-cho, Matsue, Shimane 690-8504 Japan

[13]These authors contributed equally: Prena Chaudhary, Asiri Thennakoon

✉emails: weda@mat.shimane-u.ac.jp; shlee@virginia.edu

## Abstract

The microscopic origin of the low-field 1/9 magnetization plateau in spin-1/2 kagome antiferromagnets remains unresolved. Here we show that chemical pressure reshapes the hierarchy of fractional plateaus in the titanium-based kagome family $Cs_8AB_3Ti_{12}F_{48}$ ($A$ = Rb, Li; $B$ = K, Na). High-field measurements up to 60 T reveal a robust 1/9 plateau-like phase in the expanded $Cs_8RbK_3Ti_{12}F_{48}$ and $Cs_8LiK_3Ti_{12}F_{48}$ compounds, despite the absence of the conventionally more robust 1/3 plateau. In contrast, compressed $Cs_8LiNa_3Ti_{12}F_{48}$ exhibits neither the 1/9 plateau-like phase nor a quantum-disordered ground state. Specific heat measurements and first-principles calculations show that lattice expansion preserves a frustrated, fully connected kagome exchange network and gapless quantum-disordered ground states, whereas compression reorganizes the network into weakly coupled quasi-one-dimensional subsystems and induces successive magnetic transitions. These results demonstrate that the 1/9 and 1/3 plateaus need not share a common microscopic origin and suggest that the 1/9 plateau may be a more universal feature of frustrated spin-1/2 kagome magnetism.

Suppressing classical magnetic order through geometric frustration provides a powerful route to realizing highly entangled quantum phases of matter [1–16]. Among the most extensively studied model systems are spin-1/2 Heisenberg kagome antiferromagnets (HKAFs), in which strong quantum fluctuations suppress conventional long-range magnetic order on the geometrically frustrated kagome lattice [14–16]. Theoretical studies have proposed several competing zero-field ground states for these systems, including quantum spin liquids (QSLs) with fractionalized excitations and valence-bond crystals (VBCs) [10–13,17, 18]. Under applied magnetic fields, these quantum ground states are predicted to evolve into a sequence of fractional magnetization plateaus, including a low-field 1/9 plateau followed by the more robust 1/3, 5/9, and 7/9 plateaus at higher fields [19–25].

Despite intensive theoretical and experimental efforts, this predicted hierarchy of fractional magnetization plateaus remains far from established. While the 1/3 plateau, generally regarded as the most robust fractional state, has now been observed in several kagome antiferromagnets [26–31], convincing experimental evidence for the low-field 1/9 plateau remains comparatively scarce. The conventional hierarchy of 1/3, 5/9, and 7/9 plateaus can be understood within a common picture of localized resonating magnons crystallizing on kagome hexagons into a $\sqrt{3} \times \sqrt{3}$ superstructure [21-25, 32, 33]. Remarkably, however, the 1/9 plateau is consistently predicted by numerical studies of the ideal spin-1/2 kagome Heisenberg antiferromagnet [18,21–24,34], although its microscopic nature and stabilization mechanism remain unsettled. Its recent observation in real kagome materials, where lattice distortions, exchange anisotropies, disorder, and other nonidealities are unavoidable, therefore raises a fundamental question: how robust and universal is the 1/9 plateau, and is its emergence intrinsically connected to the conventional hierarchy dominated by the 1/3 plateau? More broadly, understanding how the zero-field quantum state and material-specific perturbations control the emergence of field-induced fractional magnetization states remains an outstanding challenge in frustrated quantum magnetism.

This question becomes particularly intriguing in light of the markedly different zero-field spin correlations observed in kagome materials. In the Cu-based $YCu_3$-Br system, where a 1/9 magnetization plateau has recently been reported [27], low-energy neutron spectra are dominated by continua associated with characteristic nonzero-wavevector (q) correlations, consistent with a proposed Dirac spin-liquid state [35, 36]. By contrast, previous neutron measurements on the Ti-based kagome antiferromagnet studied here revealed strong gapless fluctuations characteristic of q = 0 correlations [37]. Whether similar field-induced fractional states can emerge from such qualitatively distinct quantum-disordered parent states remains an open question. Here we investigate the Ti-based kagome family $Cs_8AB_3Ti_{12}F_{48}$ ($A$ = Rb, Li; $B$ = K, Na), in which substitution of nonmagnetic alkali ions exerts controlled chemical pressure that systematically tunes the Ti–F–Ti superexchange network without introducing magnetic disorder. The three compounds studied here, $Cs_8LiNa_3Ti_{12}F_{48}$, $Cs_8RbK_3Ti_{12}F_{48}$, and $Cs_8LiK_3Ti_{12}F_{48}$, are hereafter denoted LiNa, RbK, and LiK, respectively.

Fig. 1 presents the crystal-structure refinements obtained from room-temperature (300 K) single-crystal neutron diffraction, confirming that the $Cs_8AB_3Ti_{12}F_{48}$ compounds ($A = \mathrm{Rb, Li}; B = \mathrm{K, Na}$) crystallize in the non-centrosymmetric monoclinic space group $Cm$. The refined lattice parameters and refinement statistics are summarized in Table 1, while comparisons of the observed and calculated structure factors are shown in Fig. 1a-c. Importantly, the refinements reveal no detectable site mixing among the constituent ions, establishing these compounds as chemically ordered realizations of the kagome $Ti^{3+}$ ($3d^1; s = 1/2$) magnetic system. This absence of detectable site mixing distinguishes the present family from many $Cu^{2+}$-based kagome compounds, in which Cu/nonmagnetic-ion site disorder can obscure the intrinsic magnetic behavior.

The $Ti^{3+}$ ions form a slightly distorted two-dimensional kagome network structurally separated by nonmagnetic alkali-ion sublattices, as shown in Fig. 1d. Each $Ti^{3+}$ ion is coordinated by six $F^-$ ligands, forming corner-sharing $TiF_6$ octahedra (Fig. 1e). The octahedra are compressed along the crystallographic [101] direction, indicated by the black arrow in Fig. 1e, producing a crystal-field splitting that stabilizes the $d_{xy}$ orbital state of the single $3d$ electron. Consequently, the magnetic orbital lies predominantly within the local kagome plane, and interactions between neighboring $Ti^{3+}$ moments are mediated primarily by Ti-F-Ti superexchange pathways.

Selective substitution at the nonmagnetic alkali-metal sites ($A$ and $B$) provides a means of tuning the chemical pressure exerted on the kagome network. As shown in Table I, incorporation of the smaller $Na^+$ ions in LiNa produces a substantially compressed unit cell volume ($V \approx 2400.5$ Å$^3$). This contraction modifies the $\mathrm{Ti} - \mathrm{F} - \mathrm{Ti}$ bond geometry and, consequently, the superexchange interactions, driving the system away from the nearly uniform, geometrically frustrated kagome limit. By contrast, the RbK and LiK variants exhibit nearly identical and comparatively expanded unit-cell volumes ($V \approx 2501.9$ Å$^3$ and $2493.8$ Å$^3$, respectively). Their crystal structures therefore remain much closer to the highly frustrated kagome geometry previously shown to host exotic field-induced fractionalized states. Full crystallographic details are provided in the Supplementary Information.

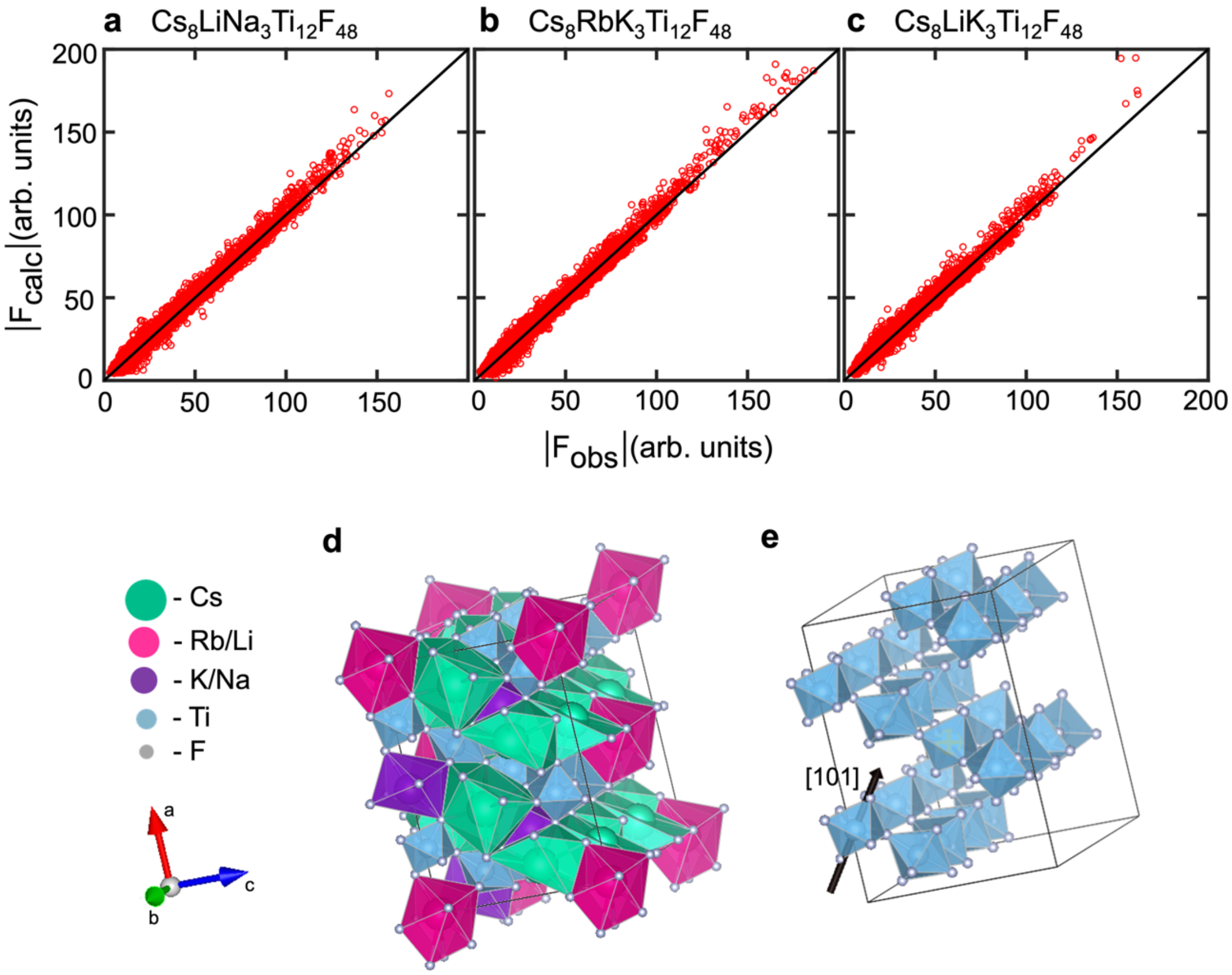

**Fig. 1 | Structural refinements and representative crystal structure of the $Cs_8AB_3Ti_{12}F_{48}$ family. a-c,** Comparison of observed and calculated structure-factor amplitudes, $|F_{calc}|$ and $|F_{obs}|$, obtained from refinements of single-crystal neutron diffraction data collected at 300 K for **a,** $Cs_8LiNa_3Ti_{12}F_{48}$, **b,** $Cs_8RbK_3Ti_{12}F_{48}$, and **c,** $Cs_8LiK_3Ti_{12}F_{48}$. The solid black line in each panel indicates ideal agreement, $|F_{calc}| = |F_{obs}|$. **d,** Polyhedral representation of the refined monoclinic crystal structure, showing the alkali-metal sublattices and $TiF_6$ octahedra. Cs, *A*-site (Rb/Li), *B*-site (K/Na), Ti, and F atoms are shown in green, magenta, purple, light blue, and gray, respectively. **e,** Ti-F framework with the alkali-metal ions omitted for clarity. In the monoclinic setting, the kagome layers are strongly inclined relative to the principal crystallographic axes and lie nearly perpendicular to the crystallographic [101] direction, indicated by the black arrow.

**Table 1 | Crystallographic space groups and room-temperature lattice parameters of the $Ti^{3+}$-based kagome fluorides.** Structural parameters for all three compounds were obtained from TOPAZ single-crystal neutron diffraction at 300 K; the RbK structural model was further refined using additional FONDER neutron and single-crystal X-ray diffraction data. The table lists the corresponding space groups, the lattice parameters *a*, *b*, and *c*, the monoclinic angle *β*, the unit-cell volume, and the refinement residuals *R* and *wR*. The numbers in parentheses represent the standard uncertainties of the last significant digits.

| 300K | $Cs_8LiNa_3Ti_{12}F_{48}$ | $Cs_8RbK_3Ti_{12}F_{48}$ | $Cs_8LiK_3Ti_{12}F_{48}$ |
|---|---|---|---|
| Space group | *Cm* | *Cm* | *Cm* |
| $a$ (Å) | 14.7933(15) | 15.151(3) | 15.253(4) |
| $b$ (Å) | 15.371(3) | 15.238(3) | 15.229(2) |
| $c$ (Å) | 10.5603(12) | 10.837(4) | 10.737(2) |
| $\alpha$ (°) | 90.0 | 90.0 | 90.0 |
| $\beta$ (°) | 91.442(10) | 90.44(2) | 90.83(2) |
| $\gamma$ (°) | 90.0 | 90.0 | 90.0 |
| $V$ (Å$^3$) | 2400.5(5) | 2501.9(10) | 2493.8(9) |
| $Z$ | 2 | 2 | 2 |
| $R$ | 0.0528 | 0.070 | 0.0643 |
| $wR$ | 0.1043 | 0.1264 | 0.1193 |

**High-field magnetization and the 1/9 plateau**

To investigate how this structural tuning modifies the underlying geometric frustration and the field-induced magnetic states, we performed high-field magnetization measurements up to 60 T. Fig. 2 shows the isothermal magnetization, $M(H)$ (top panels), and its field derivatives, d$M$/d$H$ (bottom panels), measured at 0.6 K for fields applied both perpendicular to ($H \perp$ kagome plane) and within ($H \parallel$ kagome plane) the kagome plane.

We first establish the baseline magnetic response of the compressed LiNa compound. As shown in Fig. 2a, following a rapid initial rise below 3 T, $M(H)$ increases monotonically up to 60 T for both field orientations. A pronounced magnetic anisotropy persists throughout the measured field range, with the magnetization for $H \perp$ kagome plane consistently exceeding that for $H \parallel$ kagome plane. Although the two curves remain close at low fields, their separation becomes substantially more pronounced above approximately 18 T.

Importantly, the magnetization evolves smoothly through the values corresponding to 1/9 and 1/3 of the $Ti^{3+}$ saturation moment, with no evidence of a plateau for either field orientation. This absence is further supported by the corresponding d$M$/d$H$ curves, which remain relatively featureless at high fields and exhibit no distinct minima associated with plateau boundaries.

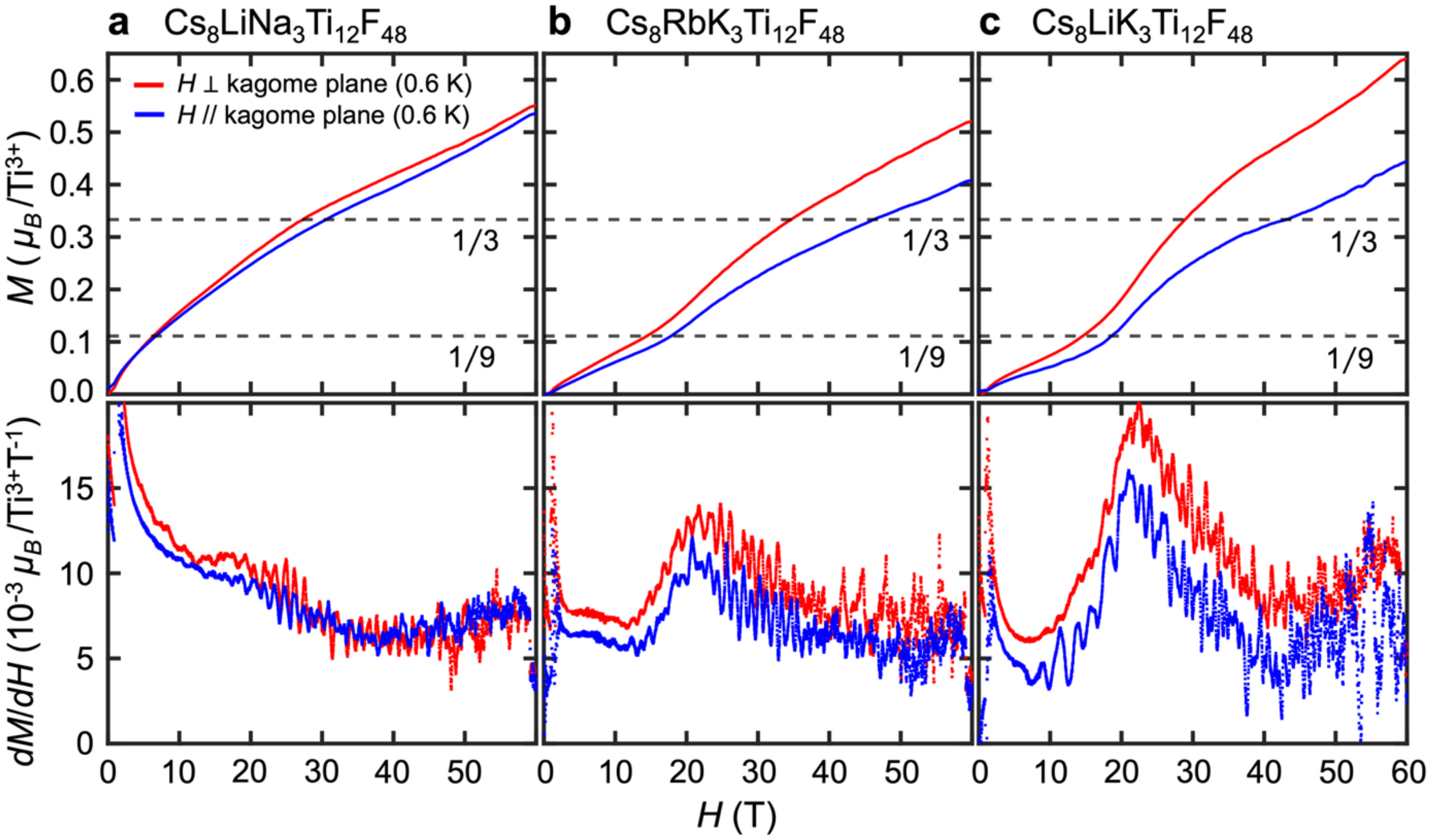

**Fig. 2 | Field-dependent magnetization and its field derivative for the $Cs_8AB_3Ti_{12}F_{48}$ family.** Field-dependent magnetization $M(H)$ (top panels) and corresponding derivatives d$M$/d$H$ (bottom panels) measured at a base temperature of 0.6 K for **a,** $Cs_8LiNa_3Ti_{12}F_{48}$, **b,** $Cs_8RbK_3Ti_{12}F_{48}$, and **c,** $Cs_8LiK_3Ti_{12}F_{48}$. Data are shown for magnetic fields applied perpendicular to the kagome plane ($H \perp$ kagome plane, red lines) and within the kagome plane ($H \parallel$ kagome plane, blue lines). The dashed horizontal lines in the top panels indicate the 1/9 and 1/3 fractional magnetization levels. Plateau-like features are absent in the structurally compressed LiNa compound, whereas pronounced 1/9 plateau-like phases emerge for the RbK and LiK, bounded by distinct minima in d$M$/d$H$.

In striking contrast, the expanded RbK and LiK compounds (Figs. 2b and 2c) exhibit clear signatures of a fractional magnetization state. Like compressed LiNa, both retain pronounced magnetic anisotropy, with the out-of-plane magnetization ($H \perp$ kagome plane) consistently exceeding the in-plane response ($H \parallel$ kagome plane) throughout the measured field range. Following a steep low-field increase, a distinct plateau-like feature emerges in the $M(H)$ curves. Analysis of the corresponding d$M$/d$H$ data identifies broad plateau-like field ranges of 2.5 T $\lesssim H \lesssim$ 15.2 T for RbK and 2.5 T $\lesssim H \lesssim$ 16.6 T for LiK, with comparable boundaries for the two field orientations.

Using exchange scale of $J/k_B \approx 47.4$ K for RbK [37] and $J/k_B \approx 40.4$ K for LiK [Supplementary Note 1], state-of-the-art numerical calculations, including density matrix renormalization group and tensor network methods, predict a 1/9 plateau over the normalized field range $0.35 < g\mu_B H/J < 0.42$ [18, 21, 34]. With $g$ factors determined by electron spin resonance measurements at 0.5 K, this corresponds to approximately 12.9-15.5 T for RbK and 10.7-12.8 T for LiK [Supplementary Note 2]. These predicted field windows lie within the broader experimentally determined plateau-like ranges, supporting the identification of the observed low-field features as 1/9 plateau-like states in both compounds.

At higher fields, the magnetization of both RbK and LiK increases rapidly through and beyond the nominal 1/3 magnetization level. The same numerical models [18, 21] predict the canonical 1/3 plateau over $0.9 < g\mu_B H/J < 1.4$, corresponding to approximately 33.2-51.6 T for RbK and 27.4-42.6 T for LiK. However, the high-field d$M$/d$H$ curves vary smoothly and show no well-defined minima up to 60 T. These results indicate that the canonical 1/3 plateau is either not stabilized in these compounds or is too weakly developed to be experimentally resolved.

Finally, temperature-dependent measurements up to 3.8 K for $H \perp$ kagome plane show that the 1/9-plateau-like minima in the expanded RbK and LiK compounds remain clearly visible at elevated temperatures [ Supplementary Note 3]. The persistence of these features indicates that the plateau-like states remain thermally robust up to at least 3.8 K. Together with the suppression of the low-temperature magnetic specific heat at 8 and 12 T in RbK and LiK, discussed below, this behavior is consistent with a depletion of low-energy magnetic excitations in the plateau-like regime.

Taken together, these high-field measurements show that the expanded RbK and LiK compounds host closely related low-temperature field-induced states, whereas chemical compression in LiNa leads to qualitatively distinct magnetic behavior.

**Ultra-Low-Temperature Specific Heat**

To further investigate the distinct low-temperature magnetic states of the compressed and expanded systems, we performed ultra-low-temperature specific-heat measurements on all three compounds, extending down to approximately 0.1 K for LiNa and 0.4 K for RbK and LiK. The phonon contribution was estimated from the specific heat of the nonmagnetic analogue $Cs_2KGa_3F_{12}$ and subtracted from the measured total specific heat to obtain the magnetic contribution, $C_\mathrm{m}$. This thermodynamic analysis provides insight into low-energy magnetic excitations and helps characterize the distinct low-temperature magnetic states in the compressed and expanded kagome antiferromagnets.

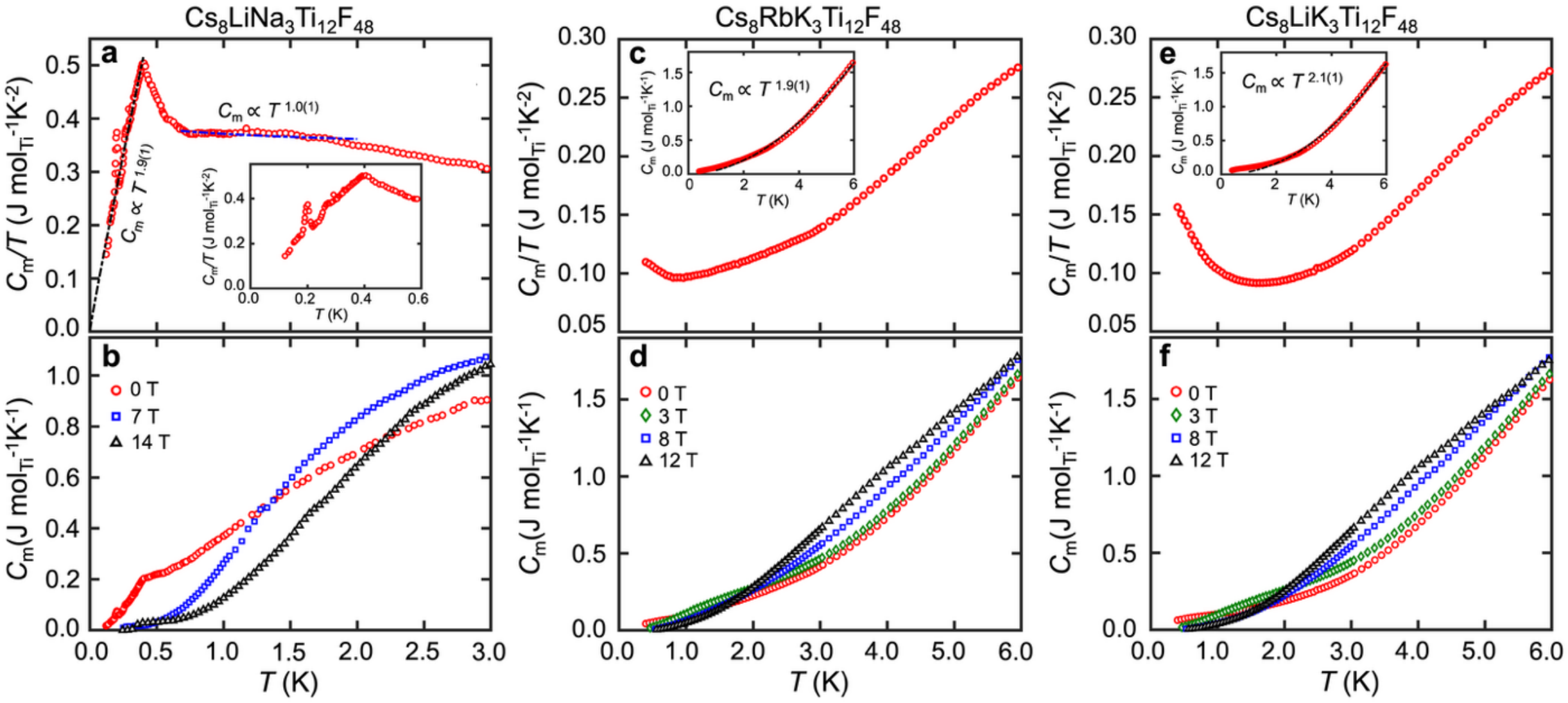

**Fig. 3 | Low-temperature magnetic specific heat of the $Cs_8AB_3Ti_{12}F_{48}$ family. a,** Zero-field magnetic specific-heat coefficient, $C_\mathrm{m}/T$, of $Cs_8LiNa_3Ti_{12}F_{48}$. The black and blue lines show fits to $C_\mathrm{m} \propto T^{1.9(1)}$ in the low-temperature region and $C_\mathrm{m} \propto T^{1.0(1)}$ above the low-temperature anomalies, respectively. The inset highlights the low-temperature anomalies. **b**, Temperature dependence of $C_\mathrm{m}$ for $Cs_8LiNa_3Ti_{12}F_{48}$ measured at 0, 7, and 14 T. **c**, Zero-field $C_\mathrm{m}/T$ of $Cs_8RbK_3Ti_{12}F_{48}$. The inset shows $C_\mathrm{m}$ together with a power-law fit, $C_\mathrm{m} \propto T^{1.9(1)}$. **d**, Temperature dependence of $C_\mathrm{m}$ for $Cs_8RbK_3Ti_{12}F_{48}$ measured at 0, 3, 8, and 12 T. **e**, Zero-field $C_\mathrm{m}/T$ of $Cs_8LiK_3Ti_{12}F_{48}$. The inset shows $C_\mathrm{m}$ together with a power-law fit, $C_\mathrm{m} \propto T^{2.1(1)}$. **f**, Temperature dependence of $C_\mathrm{m}$ for $Cs_8LiK_3Ti_{12}F_{48}$ measured at 0, 3, 8, and 12 T. All magnetic fields were applied perpendicular to the kagome plane. The numbers in parentheses indicate the uncertainty in the last quoted digit of the fitted power-law exponent.

Figs. 3a and 3b show the magnetic specific heat, $C_\mathrm{m}$, of LiNa measured in zero and applied magnetic fields for $H \perp$ kagome plane. In zero field, $C_\mathrm{m}$ exhibits distinct anomalies near 0.26 and 0.40 K (the inset of Fig. 3a), suggesting two successive magnetic transitions, whereas the sharp peak near 0.20 K is attributed to an experimental artifact. Above 0.4 K, $C_\mathrm{m}/T$ remains nearly constant, corresponding to an approximately linear temperature dependence of $C_\mathrm{m} \propto T^{1.0(1)}$ over the range 0.68–2 K. Such behavior is consistent with dominant quasi-one-dimensional magnetic correlations. By contrast, below 0.4 K, $C_{\mathrm{m}}/T$ varies approximately linearly with temperature, corresponding to $C_{\mathrm{m}} \propto T^{1.9(1)}$ over the range 0.12–0.4 K. The crossover from nearly linear to quadratic temperature dependence suggests that couplings between the quasi-one-dimensional magnetic units become increasingly important below 0.4 K, giving rise to more two-dimensional magnetic correlations. Whether this low-temperature state develops long-range magnetic order remains an open question requiring further investigation.

Fig. 3b shows that the anomalies at 0.26 and 0.4 K are rapidly suppressed by an applied magnetic field, supporting their magnetic origin. At 7 and 14 T, $C_{\mathrm{m}}$ decreases rapidly toward zero upon cooling, indicating a strong depletion of low-energy magnetic excitations. This behavior is consistent with the opening of a field-induced excitation gap.

By contrast, RbK and LiK exhibit no sharp specific-heat anomaly down to approximately 0.4 K (Figs. 3c–f). In zero-field, the magnetic specific heat follows a power-law temperature dependence, $C_{\mathrm{m}} \propto T^{n}$, with $n = 1.9(1)$ and 2.1(1) for RbK and LiK, respectively, over the range 1–6 K. This approximately $T^2$ behavior contrasts markedly with the nearly linear $C_{\mathrm{m}}(T)$ observed in LiNa and is consistent with two-dimensional low-energy magnetic excitations, although the power-law dependence alone does not uniquely establish their microscopic origin.

The effect of magnetic field on $C_{\mathrm{m}}$ in RbK and LiK also varies strongly with temperature (Figs. 3d and f). At the lowest temperatures, applied fields of 8 and 12 T suppress $C_{\mathrm{m}}$ relative to the zero-field response, indicating a depletion of low-energy magnetic excitations. With increasing temperature, the field-dependent curves cross the zero-field data, and $C_{\mathrm{m}}$ becomes enhanced at higher fields above approximately 1.8 K. This behavior is consistent with the development of a field-induced gap at low energies accompanied by a shift of magnetic excitations to higher energies. No sharp field-induced thermodynamic anomaly is observed within the measured temperature range. The closely similar field evolution in RbK and LiK is consistent with their related high-field magnetization behavior and the stabilization of the 1/9 plateau state in both compounds.

Together, these results establish a clear thermodynamic distinction between the larger-volume RbK and LiK compounds and the structurally compressed LiNa compound. Whereas LiNa develops low-temperature magnetic transitions that are rapidly suppressed by an applied field, RbK and LiK exhibit no sharp thermodynamic anomalies within the measured temperature range and display closely similar power-law and field-dependent specific-heat behavior. Their common depletion of low-energy magnetic excitations at high fields further reinforces the close correspondence between the low-temperature magnetic states of RbK and LiK and their field-induced 1/9-plateau-like behavior.

**First-principles exchange networks and chemical control of frustration**

To identify the microscopic origin of the contrasting low-temperature magnetic states, we performed DFT calculations for RbK, LiK, and LiNa using the neutron-refined crystal structures. Fig. 4 shows that, although all three compounds share the same monoclinic structure with 14 symmetry-inequivalent nearest-neighbor exchange pathways, their exchange hierarchies differ markedly.

For LiNa, 40 exchange interactions were evaluated. Both interlayer and in-plane second-neighbor couplings are much weaker than the dominant nearest-neighbor interactions, allowing the Hamiltonian to be reduced to the 14 symmetry-inequivalent nearest-neighbor exchanges (Supplementary Table 7; Figs. 4a and 4b). Strikingly, the LiNa exchange network separates into two weakly coupled quasi-one-dimensional magnetic subsystems, denoted chain-A and chain-B. Chain-B contains three strong antiferromagnetic interactions, including the strongest nearest-neighbor coupling, $J_{1b\text{-}2b}$ = 80.5 K, and is connected to the rest of the network by much weaker zigzag couplings, including two weak ferromagnetic interactions. The remaining seven nearest-neighbor interactions form the chain-A subsystem, including two of the strongest couplings, $J_{2c\text{-}2c}$ = 75.3 K and $J_{3b\text{-}3b}$ = 74.7 K, together with five moderate-to-strong antiferromagnetic couplings that form an anisotropic zigzag-chain network.

For RbK and LiK, longer-range interactions up to the 31$^{st}$-neighbor are negligible within the computational uncertainty, and their magnetic Hamiltonians are likewise well described by the 14 nearest-neighbor exchanges (Supplementary Table 7; Figs. 4a, 4c, and 4d). In sharp contrast to LiNa, however, all 14 nearest-neighbor interactions are antiferromagnetic in both compounds, preserving fully connected and frustrated kagome exchange networks despite substantial variations in the individual exchange strengths.

Notably, these pronounced differences in exchange hierarchy occur despite comparable nearest-neighbor Ti-Ti distances across the three compounds, demonstrating that the superexchange interactions are highly sensitive to the detailed Ti-F-Ti geometry and local chemical environment. Alkali-metal substitution therefore does more than simply renormalize the exchange energy scale: it qualitatively reorganizes the magnetic Hamiltonian, transforming the weakly coupled quasi-one-dimensional subsystems of LiNa into fully connected frustrated kagome networks in RbK and LiK.

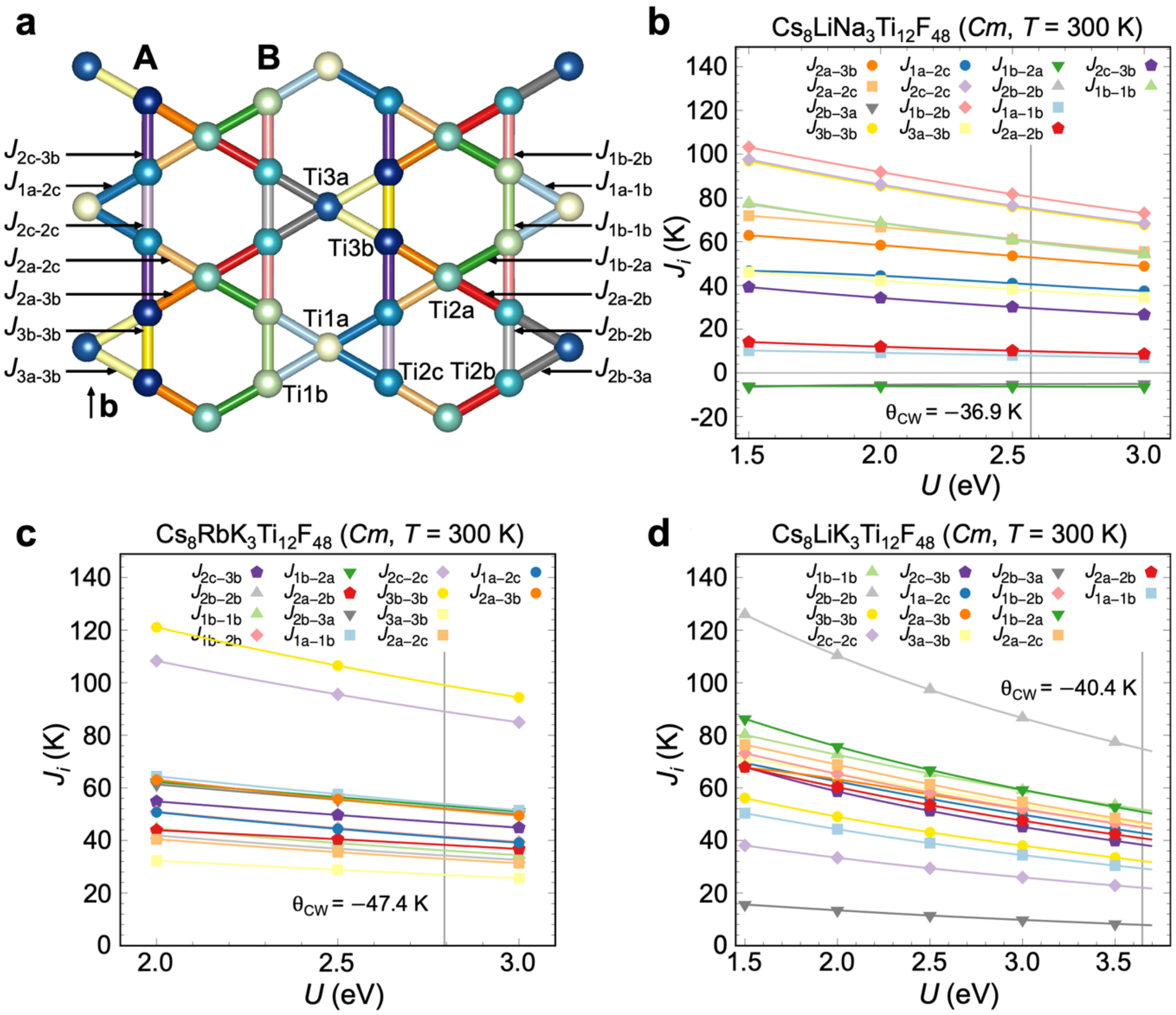

**Fig. 4 | DFT-derived nearest-neighbor exchange interactions and exchange networks of the $Cs_8AB_3Ti_{12}F_{48}$ family. a,** Common nearest-neighbor exchange network of the $Cs_8AB_3Ti_{12}F_{48}$ family, showing the 14 symmetry-inequivalent exchange pathways and their division into A and B exchange subsystems. Both subsystems extend along the crystallographic *b* direction and provide the structural basis for the chain-A and chain-B magnetic subsystems that become weakly coupled and quasi-one-dimensional in $Cs_8LiNa_3Ti_{12}F_{48}$. **b-d,** Fourteen symmetry-inequivalent nearest-neighbor exchange interactions calculated as a function of the on-site Coulomb interaction *U* using the 300 K *Cm* crystal structures of **b,** $Cs_8LiNa_3Ti_{12}F_{48}$, **c,** $Cs_8RbK_3Ti_{12}F_{48}$, and **d,** $Cs_8LiK_3Ti_{12}F_{48}$. Vertical lines indicate the values of *U* selected to reproduce the experimental Curie-Weiss temperatures, $\theta_{CW} = -36.9$ K for $Cs_8LiNa_3Ti_{12}F_{48}$, $-47.4$ K for $Cs_8RbK_3Ti_{12}F_{48}$, and $-40.4$ K for $Cs_8LiK_3Ti_{12}F_{48}$.

The central finding of this work is that chemical tuning separates the low-field 1/9 magnetization plateau from the conventional hierarchy of fractional plateaus predicted for the spin-1/2 Heisenberg kagome antiferromagnet. In both expanded RbK and LiK, a robust 1/9 plateau-like phase emerges under magnetic field, whereas no evidence for the conventionally robust 1/3 plateau is observed up to 60 T. This contrasts with the standard theoretical hierarchy, in which the 1/3 plateau is dominant and the 1/9 plateau considerably more delicate. Our results demonstrate that these plateaus need not emerge together and suggest that the 1/9 plateau is a distinct field-induced state whose stability is largely independent of the conventional higher-field sequence.

The higher-field 1/3, 5/9, and 7/9 plateaus are commonly understood within a localized-magnon framework. The flat magnon band of the ideal Heisenberg kagome antiferromagnet permits excitations to localize on hexagonal plaquettes, which crystallize at commensurate magnetizations into a $\sqrt{3} \times \sqrt{3}$ superstructure, producing the canonical plateau sequence [25]. These states can therefore be viewed as different fillings of closely related resonating-hexagon crystals. The microscopic origin of the 1/9 plateau, however, remains unsettled, with proposals ranging from enlarged valence-bond crystals and symmetry-broken phases to topological descendants of quantum spin liquids. Its persistence in the complete absence of the 1/3 plateau strongly suggests that the 1/9 state is not simply the low-field member or precursor of the conventional localized-magnon hierarchy.

Comparison with other kagome materials provides a broader perspective. In the Cu-based $YCu_3$-Br system, the 1/9 plateau emerges from a quantum-disordered state characterized by nonzero-wavevector continua associated with a proposed Dirac spin-liquid and prominent spectral weight at q = (1/3, 0) [27,35,36]. In contrast, neutron scattering on Ti-based RbK reveals strong gapless q = 0 fluctuations and a broad dispersive continuum [37]. The appearance of a similar 1/9 fractional state from such different zero-field correlations suggests that the 1/9 plateau is not uniquely tied to a particular parent spin-liquid state or ordering wavevector. Instead, the 1/9 plateau may represent a robust field-induced quantum state accessible from a broader class of frustrated kagome ground states.

Chemical substitution within the $Cs_8AB_3Ti_{12}F_{48}$ family provides a controlled means of testing the conditions under which this state survives. DFT calculations show that chemical pressure primarily reorganizes the nearest-neighbor exchange hierarchy rather than introducing substantial longer-range interactions. In the expanded RbK and LiK compounds, all 14 nearest-neighbor interactions are antiferromagnetic, preserving fully connected, frustrated kagome exchange networks despite significant bond inequivalence. In compressed LiNa, by contrast, the kagome network decomposes into two weakly coupled quasi-one-dimensional magnetic subsystems, partially relieving geometric frustration. This evolution provides a clean route for disentangling the role of exchange connectivity, quantum disorder, and field-induced fractionalization.

The thermodynamic measurements reinforce this connection. The expanded compounds exhibiting the 1/9 plateau-like phase remain strongly fluctuating to the lowest measured temperatures, with no sharp specific-heat anomaly and an approximately quadratic temperature dependence, consistent with gapless low-energy excitations of a quantum-disordered state. Previous inelastic neutron-scattering on RbK reveals a broad magnetic continuum [37]. In contrast, compressed LiNa exhibits successive low-temperature specific-heat anomalies and a crossover from nearly linear to quadratic temperature dependence, consistent with the development of quasi-one-dimensional correlations followed by low-temperature magnetic transitions and higher-dimensional correlations. The DFT-derived exchange hierarchy and recent neutron-scattering measurements further support this picture, indicating that compression relieves frustration by driving LiNa toward weakly coupled spin-1/2 "chainsaw" subsystems [38]. Together, these observations establish a strong correlation between a quantum-disordered kagome state and stabilization of the low-field 1/9 plateau-like phase, while showing that the momentum-space structure of the parent quantum state need not be universal.

Although our experiments do not determine the microscopic wavefunction of the 1/9 state, they impose two important constraints on theory. First, the 1/9 plateau must survive even when the conventionally robust 1/3 plateau is absent, arguing against a simple common plateau hierarchy. Second, it must be compatible with markedly different zero-field correlations, from the nonzero-wavevector continuum of Cu-based kagome systems to the dominant $q = 0$ fluctuations of the Ti-based RbK. These results point to the low-field 1/9 plateau as a distinct and unusually robust field-induced quantum state rather than a subordinate member of the localized-magnon sequence. More broadly, the $Cs_8AB_3Ti_{12}F_{48}$ family demonstrates how controlled chemical tuning can reveal the organizing principles of fractional quantum states and suggests that the 1/9 plateau may be a more universal feature of frustrated spin-1/2 kagome magnetism than previously recognized.

## METHODS

### Sample Synthesis and Structural Characterization

High-quality single crystals of the $Ti^{3+}$-based kagome fluorides RbK, LiK, and LiNa were grown by a high-temperature alkali-chloride flux method. Dried starting materials and purified $TiF_3$ precursors were mixed stoichiometrically in nickel crucibles, heated to 800 °C under argon, and slowly cooled at 2 °C/h. The residual flux was dissolved in deionized water. The resulting highly insulating, dark-brown crystals exhibited well-defined facets and typical dimensions of ~ 4 × 4 × 1 $mm^3$.

The crystal structures of all three compounds were determined at 300 K using single-crystal neutron diffraction on the TOPAZ time-of-flight diffractometer at the Spallation Neutron Source [39]. The data were reduced using Mantid and refined in the monoclinic space group $Cm$ using Jana2020 [40]. For RbK, additional room-temperature neutron diffraction data were collected using the FONDER four-circle diffractometer at JRR-3 [41] and jointly refined with in-house single-crystal X-ray diffraction data. This high-precision structural model was used for the representative structure in Fig. 1d. Further refinement details are provided in the Supplementary Information.

### High-Field Magnetization Measurements

Preliminary high-field magnetization measurements (Supplementary Note 7) were performed on powder samples at the International MegaGauss Science Laboratory, Institute for Solid State Physics, University of Tokyo. Pulsed-field magnetization, $M(H)$, was subsequently measured up to 60 T and down to 0.6 K at the National High Magnetic Field Laboratory (NHMFL), Los Alamos National Laboratory, using an induction method with coaxial pickup coils. Magnetic anisotropy was probed with fields applied parallel and perpendicular to the kagome layers. The pulsed-field data were calibrated against low-field magnetization measured up to 9 T using a commercial SQUID-based Magnetic Property Measurement System (MPMS) at the University of Virginia.

## Electron Spin Resonance Measurements

High-field electron spin resonance (ESR) measurements were performed on single crystals of LiNa, RbK, and LiK at the Institute for Materials Research, Tohoku University, using the high-field THz-ESR (TESLA-ESR) system equipped with a $^{3}$He ultralow-temperature cryostat and pulsed magnetic fields. Frequency-field measurements were conducted at 0.5 K with the $H \perp$ kagome plane. Resonance fields were determined from the dominant absorption minima, and the frequency-field dispersion was fitted linearly to extract the effective $g$ factors. Further details are provided in the Supplementary Information.

## Specific Heat Measurements

Low-temperature specific-heat measurements were performed to probe low-energy magnetic excitations and possible thermodynamic phase transitions. The magnetic contribution, $C_{\mathrm{m}}$, was obtained by subtracting the phonon contribution, estimated from the nonmagnetic analogue $Cs_2KGa_3F_{12}$, after appropriate normalization and interpolation to the corresponding sample temperatures. The nonmagnetic reference was measured down to approximately 0.4 K. Below this temperature, the phonon contribution was extrapolated using the low-temperature Debye form, $C_{\mathrm{ph}} = \beta T^3$, where $\beta$ was determined from a fit to the lowest-temperature $Cs_2KGa_3F_{12}$ data. Measurements were conducted using the thermal-relaxation method in Physical Property Measurement Systems (PPMS, Quantum Design) equipped with dilution-refrigerator (DR) or $^{3}$He inserts at Johns Hopkins University and Fudan University. RbK and LiK were measured down to approximately 0.4 K, whereas measurements on LiNa were extended to approximately 0.1 K to resolve the low-temperature thermodynamic anomalies near 0.4 K.

## Density-functional-theory calculations and exchange-parameter determination

Magnetic exchange interactions in RbK, LiK and LiNa were calculated using FPLO [42] within the generalized-gradient approximation [43], with $Ti^{3+}$ correlations treated using DFT+$U$ approach [44]. The interactions were mapped into a Heisenberg Hamiltonian

$$H = \sum_{i<j} J_{ij} \mathbf{S}_i . \mathbf{S}_j, \qquad (1)$$

where $S = 1/2$; positive and negative $J_{ij}$ denote antiferromagnetic and ferromagnetic exchange interactions, respectively.

Exchange parameters were obtained by mapping the total energies of multiple collinear spin configurations onto the corresponding classical Heisenberg energies, following Refs. [45–47]. The Hund coupling $J_{\mathrm{H}}$ was fixed at the literature value [48], while the on-site Coulomb interaction $U$ was varied. For each compound, $U$ was selected by matching the calculated Curie-Weiss temperature to the experimental value; the selected values are indicated by the vertical lines in Figs. 4b-d for LiNa, RbK, and LiK, respectively.

Longer-range interactions were also evaluated to test the validity of a nearest-neighbor description. For RbK and LiK, interactions up to the 31$^{st}$ neighbor were calculated, while 40 exchange interactions were evaluated for LiNa.In all three compounds, the longer-range couplings were much weaker than the dominant nearest-neighbor interactions, supporting magnetic Hamiltonians described primarily by the 14 symmetry-inequivalent nearest-neighbor exchanges shown in Fig. 4.

## ACKNOWLEDGEMENTS

We thank Collin Broholm for helpful discussions and Jong K. Keum for assistance with Laue diffraction measurements performed at the SNS X-Ray Laboratory.

## AUTHOR CONTRIBUTIONS

H.U. and S.-H.L. conceived and designed the research. H.U. grew the single crystals. P.C., T.P., C.H., and J.H.H. performed single-crystal neutron diffraction measurements on all three compounds at the SNS, ORNL. P.C., A.T., T.M., Y.N., and T.J.S. performed single-crystal neutron diffraction measurements on RbK at the JRR-3 reactor. P.C., A.T., L.W., and N.H. carried out high-field magnetization measurements on single crystals, while A.M., K.K., and H.U. performed preliminary high-field magnetization measurements on powder samples. H.W., L.Z., A.T., P.C., and S.L. performed specific heat measurements. T.P. and H.N. performed electron spin resonance measurements. H.O.J. carried out density-functional-theory calculations.

All authors except H.O.J. contributed to the analysis of the high-field magnetization, specific heat, electron spin resonance, and neutron scattering data. P.C., A.T., T.P., H.O.J., H.U., G.-W.C., and S.-H.L. prepared the initial draft of the manuscript, and all authors contributed to its revision and approved the final version.

## FUNDING

P.C., A.T., T.P., G.W.C., and S.-H.L. were supported by the U.S. Department of Energy, Office of Science, Basic Energy Sciences, under Award No. DE-SC0026087 for the project "Quantum spin states of new Ti-based kagome antiferromagnets." H.U. was supported by JSPS KAKENHI Grant No. 26K00656. L.Z. was supported by the U.S. Department of Energy, Office of Science, Basic Energy Sciences, under Award No. DE-SC0024469. Work performed at Fudan University was supported by the National Science Foundation of China under Grant No. 125B2074. The National High Magnetic Field Laboratory is supported by the National Science Foundation through NSF/DMR-2128556, the State of Florida, and the U.S. Department of Energy. A portion of this research used resources at the Spallation Neutron Source, a U.S. Department of Energy Office of Science User Facility operated by the Oak Ridge National Laboratory, and the FONDER four-circle diffractometer at the JRR-3 research reactor of the Japan Atomic Energy Agency. The work at Tohoku University and University of Tokyo was partly supported by the Grants-in-Aid for Scientific Research from Japan Society for the Promotion of Science (Grant Nos. JP22H00101 and 23KK0051). H.O.J. acknowledges support through JSPS KAKENHI Grant No. 25K08460.

# Supplementary Information

## for

## Universality of the 1/9 Magnetization Plateau and Quantum-Disordered States in the Kagome Family $Cs_8AB_3Ti_{12}F_{48}$ ($A$ = Rb, Li; $B$ = K, Na)

Prena Chaudhary[1], Asiri Thennakoon[1], Tommy Park[1], Hanru Wang[2], Leshan Zhao[3], Laurel Winter[4], Neil Herrison[4], Christina Hoffmann[5], Junghong H. He[5], Harald O. Jeschke[6], Hiroyuki Nojiri[7], Akira Matsuo[8], Koichi Kindo[8], Miwako Takahashi[9], Yukio Noda[8,10], Taku J. Sato[8,10,11], Shiyan Li[2], Hiroaki Ueda[12], Gia-Wei Chern[1], and Seung-Hun Lee[1]

[1]Department of Physics, University of Virginia; Charlottesville, Virginia, 22904, USA.

[2]State Key Laboratory of Surface Physics, Department of Physics, Fudan University, Shanghai 200438, China

[3]Institute for Quantum Matter and Department of Physics and Astronomy, The Johns Hopkins University; Baltimore, Maryland, 21218, USA.

[4]National High Magnetic Field Laboratory, Los Alamos National Laboratory, Los Alamos, NM, 87545, USA

[5]Oak Ridge National Laboratory, Oak Ridge, Tennessee, 37831, USA.

[6]Research Institute for Interdisciplinary Science, Okayama University, Okayama 700-8530, Japan

[7]Institute for Materials Research, Tohoku University, Sendai 980-8577, Japan

[8]Institute for Solid State Physics, University of Tokyo, Kashiwa, 277-8581, Japan

[9]Department of Materials Science, Institute of Pure and Applied Sciences, University of Tsukuba, Tsukuba, Ibaraki, 305-8573, Japan

[10]Institute of Multidisciplinary Research for Advanced Materials, Tohoku University, Sendai 980-8577, Japan

[11]Trans-scale Quantum Science Institute, University of Tokyo, Tokyo 113-0033, Japan

[12]Co-Creation Institute for Advanced Materials, Shimane University, 1060 Nishikawatsu-cho, Matsue, Shimane 690-8504 Japan

**SUPPLEMENTARY NOTE 1: Magnetic susceptibility and Curie-Weiss analysis of the $Cs_8\mathit{AB}_3Ti_{12}F_{48}$ family.**

The temperature-dependent magnetic susceptibilities of $Cs_8LiNa_3Ti_{12}F_{48}$, $Cs_8LiK_3Ti_{12}F_{48}$, and $Cs_8RbK_3Ti_{12}F_{48}$ exhibit broad behavior characteristic of antiferromagnetically correlated systems. To estimate the dominant magnetic interaction scale, the inverse susceptibility, $1/\chi(T)$, was fitted to the Curie-Weiss relation over 150-300 K, where short-range correlation effects are reduced. The fits yield Curie-Weiss temperatures of $\theta_{\mathrm{CW}} = -36.9$ K for LiNa, $\theta_{\mathrm{CW}} = -40.4$ K for LiK, and $\theta_{\mathrm{CW}} = -47.4$ K for RbK. The negative Curie-Weiss temperatures indicate dominant antiferromagnetic interactions in all three compounds.

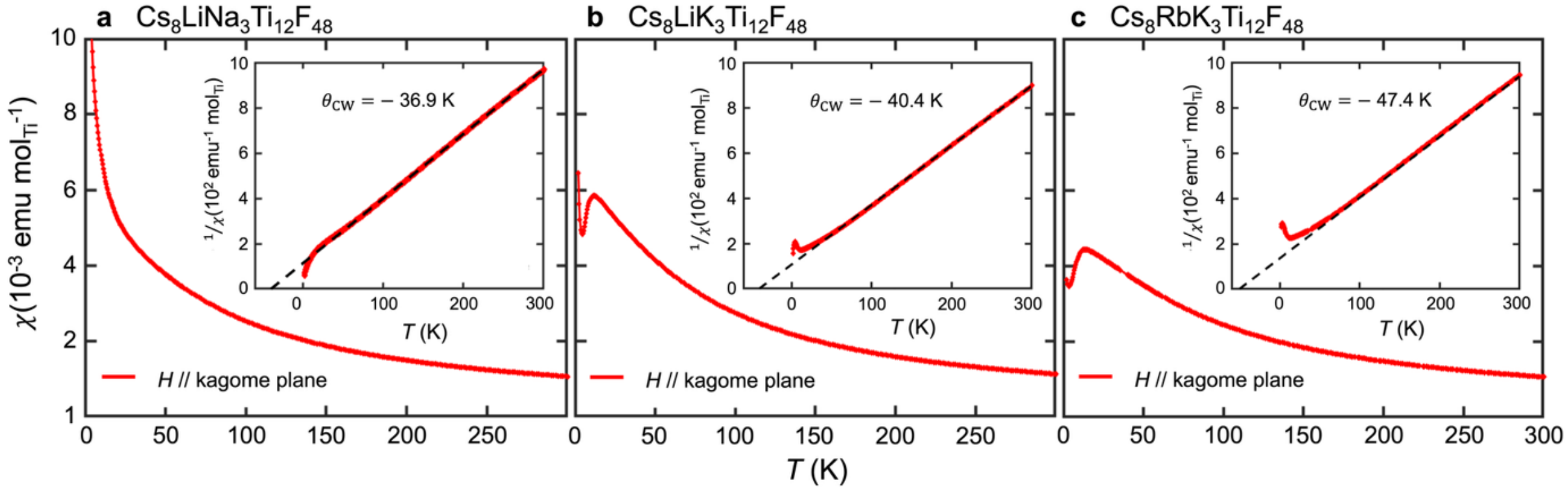

**Supplementary Fig. 1 | Temperature-dependent magnetic susceptibility and Curie–Weiss analysis of the $Cs_8\mathit{AB}_3Ti_{12}F_{48}$ family.** Single-crystal magnetic susceptibility $\chi(T)$ measured from approximately 2 to 300 K under an applied magnetic field of $H$ = 1 T parallel to the kagome plane for **a,** $Cs_8LiNa_3Ti_{12}F_{48}$; **b,** $Cs_8LiK_3Ti_{12}F_{48}$; and **c,** $Cs_8RbK_3Ti_{12}F_{48}$. Insets show the corresponding inverse susceptibilities, $1/\chi(T)$, and Curie–Weiss fits over 150–300 K.

**SUPPLEMENTARY NOTE 2: Electron Spin Resonance (ESR) measurements**

**Experimental procedure**

High-field electron spin resonance (ESR) measurements were performed in transmission geometry on single crystals of LiNa, RbK, and LiK at the Institute for Materials Research, Tohoku University. Measurements were conducted using the high-field THz-ESR (TESLA-ESR) system equipped with a $^{3}$He low-temperature cryostat and pulsed magnetic fields generated by capacitor discharge. Each crystal was mounted separately, with the microwave frequency and temperature varied between runs. The measurements relevant to the present analysis covered the frequency range 135-405 GHz. The frequency-field data used to determine the effective $g$ factors were collected at 0.5 K with the magnetic field applied perpendicular to the kagome plane ($H \perp$ kagome plane). The RbK spectrum measured at 135 GHz was excluded from the quantitative analysis because the resonance feature was strongly affected by noise and background contributions.

**ESR spectra and resonance-field determination**

Representative frequency-dependent ESR spectra measured at 0.5 K are shown in Supplementary Fig. 2. For visualization, the displayed spectra were linearly background-subtracted, smoothed using a five-point moving average, normalized to the resonance-dip amplitude, and vertically offset. These processing steps were applied only for presentation; all resonance-field fits were performed on the unprocessed raw data using the SciPy Python library.

The ESR spectra exhibit broad absorption minima spanning several tesla, which shift systematically toward higher magnetic fields with increasing microwave frequency. This behavior is clear resolved at lower frequencies but becomes less distinct at higher frequencies as the signal-to-noise ratio decreases. For each frequency, the resonance field, $B_0$, was defined as the center of the dominant absorption minimum obtained from a Lorentzian or Gaussian line-shape fit, as appropriate for the observed profile. The analysis assumed a single dominant broad absorption feature within the selected field window.

The measurement system recorded one magnetic-field-monitor channel (CH1) and three same-frequency microwave-transmission channels. Only the rising-field portion of the pulsed-field trace was used for the resonance-field analysis. The magnetic field was obtained from $B = f_{\mathrm{cal}} \times \mathrm{CH1}$, where $f_{\mathrm{cal}} = 0.881$ T V$^{-1}$ is the field-calibration factor.

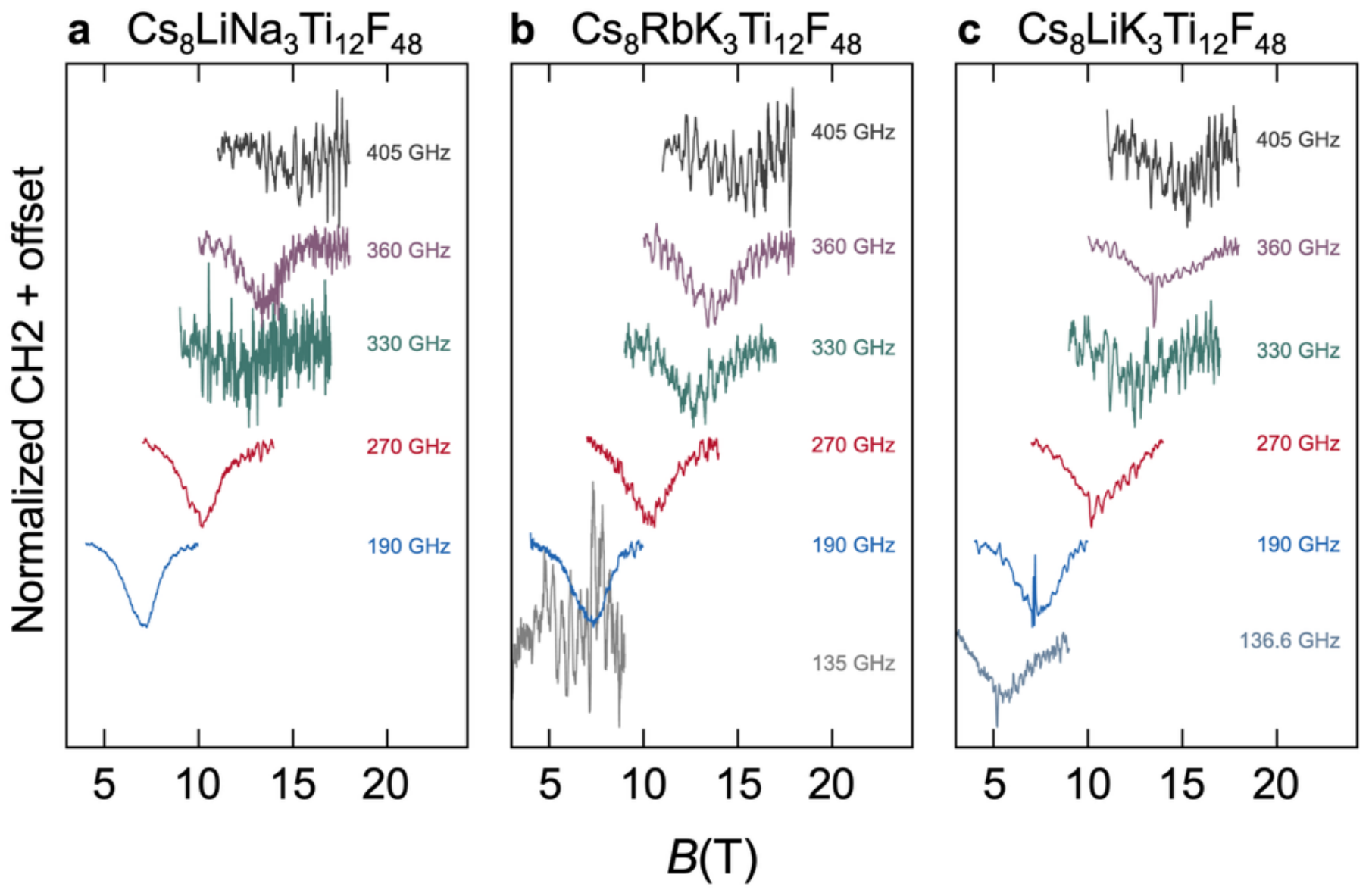

**Supplementary Fig. 2 | Frequency-dependent high-field ESR spectra measured at 0.5 K. a,** $Cs_8LiNa_3Ti_{12}F_{48}$; **b,** $Cs_8RbK_3Ti_{12}F_{48}$; and **c,** $Cs_8LiK_3Ti_{12}F_{48}$. Normalized rise-field CH2 transmission spectra are plotted as a function of magnetic field and vertically offset for clarity. The microwave frequency is indicated next to each trace. The spectra shown here were processed only for visualization, whereas the resonance fields used in the quantitative analysis were obtained by fitting the unprocessed raw data. The gray RbK spectrum measured at 135 GHz was excluded from the frequency-field analysis because the resonance feature was obscured by a low signal-to-noise ratio and substantial background contribution.

## Determination of the effective $g$ factor

The resonance frequencies were plotted as a function of the corresponding resonance field, as shown in Supplementary Fig. 3. For an ESR transition dominated by Zeeman splitting, the resonance frequency and magnetic field are related by $h\nu = g_{\mathrm{eff}}\mu_{\mathrm{B}}B_0$, where $h$ is Planck's constant, $\nu$ is the microwave frequency, $g_{\mathrm{eff}}$ is the effective spectroscopic $g$ factor, and $\mu_{\mathrm{B}}$ is the Bohr magneton. To allow for a possible small offset, the measured frequency-field dependence was fitted using $\nu = mB_0 + c$, and the effective $g$ factor was calculated from $g_{\mathrm{eff}} = \frac{m}{\mu_{\mathrm{B}}/h} = \frac{m}{13.996245\ \mathrm{GHz\ T^{-1}}}$, where $\mu_{\mathrm{B}}/h = 13.996245\ \mathrm{GHz\ T^{-1}}$ and $m$ is the fitted slope.

The accepted resonance points show an approximately linear frequency-field dependence (supplementary Fig. 3), yielding $g_{\mathrm{eff}} = 1.940(30)$ for LiNa, $g_{\mathrm{eff}} = 1.915(33)$ for RbK, and $g_{\mathrm{eff}} = 1.975(70)$ for LiK. The resonance-center uncertainties were propagated into the weighted regression. The quoted uncertainties are the standard errors of the fitted slopes and do not include systematic uncertainty in the magnetic field at the sample.

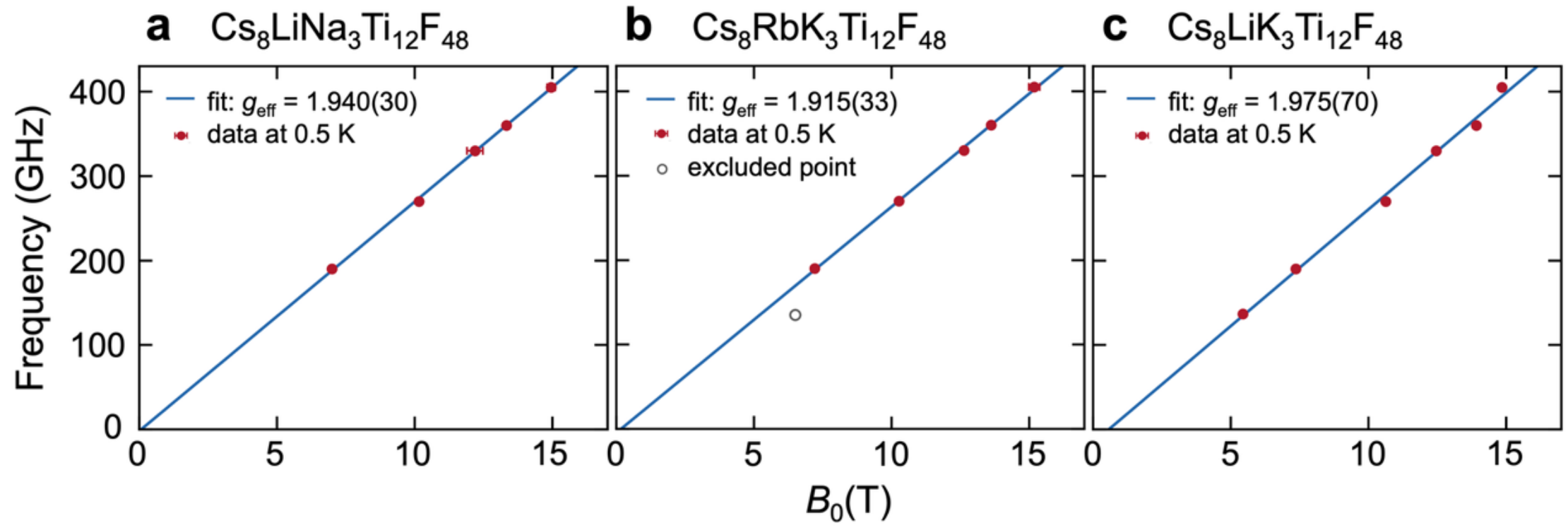

**Supplementary Fig. 3 | Frequency-field dispersion at 0.5 K. a,** $Cs_8LiNa_3Ti_{12}F_{48}$; **b,** $Cs_8RbK_3Ti_{12}F_{48}$; and **c,** $Cs_8LiK_3Ti_{12}F_{48}$. Filled red symbols show the resonance fields included in the analysis, and solid lines represent linear fits of $\nu = mB_0 + c$, with the intercept allowed to vary. The open symbol for RbK at 135 GHz was excluded because of its low signal-to-noise ratio and background contributions. Horizontal error bars show the fitted uncertainties in $B_0$.

**Estimated field range of the 1/9 magnetization plateau**

To relate the ESR-derived $g$ factors to the high-field magnetization measurements, the theoretical field interval associated with the 1/9 magnetization plateau, $0.35 < g\mu_B H/J < 0.42$, was converted into magnetic field units. For each boundary of this interval, $x = 0.35$ and $0.42$, the corresponding magnetic field is $H = x\left(\frac{k_B}{\mu_B}\right)\frac{J/k_B}{g_{\text{eff}}}$. The characteristic exchange scale was estimated from the magnitude of the Curie-Weiss temperature, using $J/k_B \approx |\theta_{\text{CW}}|$. Using $|\theta_{\text{CW}}| = 47.4$ K for RbK and 40.4 K for LiK gives the field ranges summarized in Supplementary Table 1. LiNa is not included in this comparison because no clear 1/9 magnetization plateau-like feature is observed in its high-field magnetization.

**Supplementary Table 1 | Theoretical 1/9 plateau field ranges calculated using the ESR-derived effective $g$ values at 0.5 K.** The field ranges were obtained from the theoretical dimensionless interval $0.35 < g\mu_B H/J < 0.42$ using $J/k_B \approx |\theta_{\text{CW}}|$ as an estimate of the characteristic exchange scale. The central values of the ESR-derived $g_{\text{eff}}$ factors were used to calculate the lower and upper field boundaries.

| Compound | $g_{\text{eff}}$ at 0.5 K | $J/k_B$(K) | Estimated 1/9 plateau frange (T) |
|---|---|---|---|
| RbK | 1.915(33) | 47.4 | 12.9 – 15.5 |
| LiK | 1.975(70) | 40.4 | 10.7 – 12.8 |

**SUPPLEMENTARY NOTE 3: Temperature dependence of highfield magnetization for fields perpendicular to the kagome plane.**

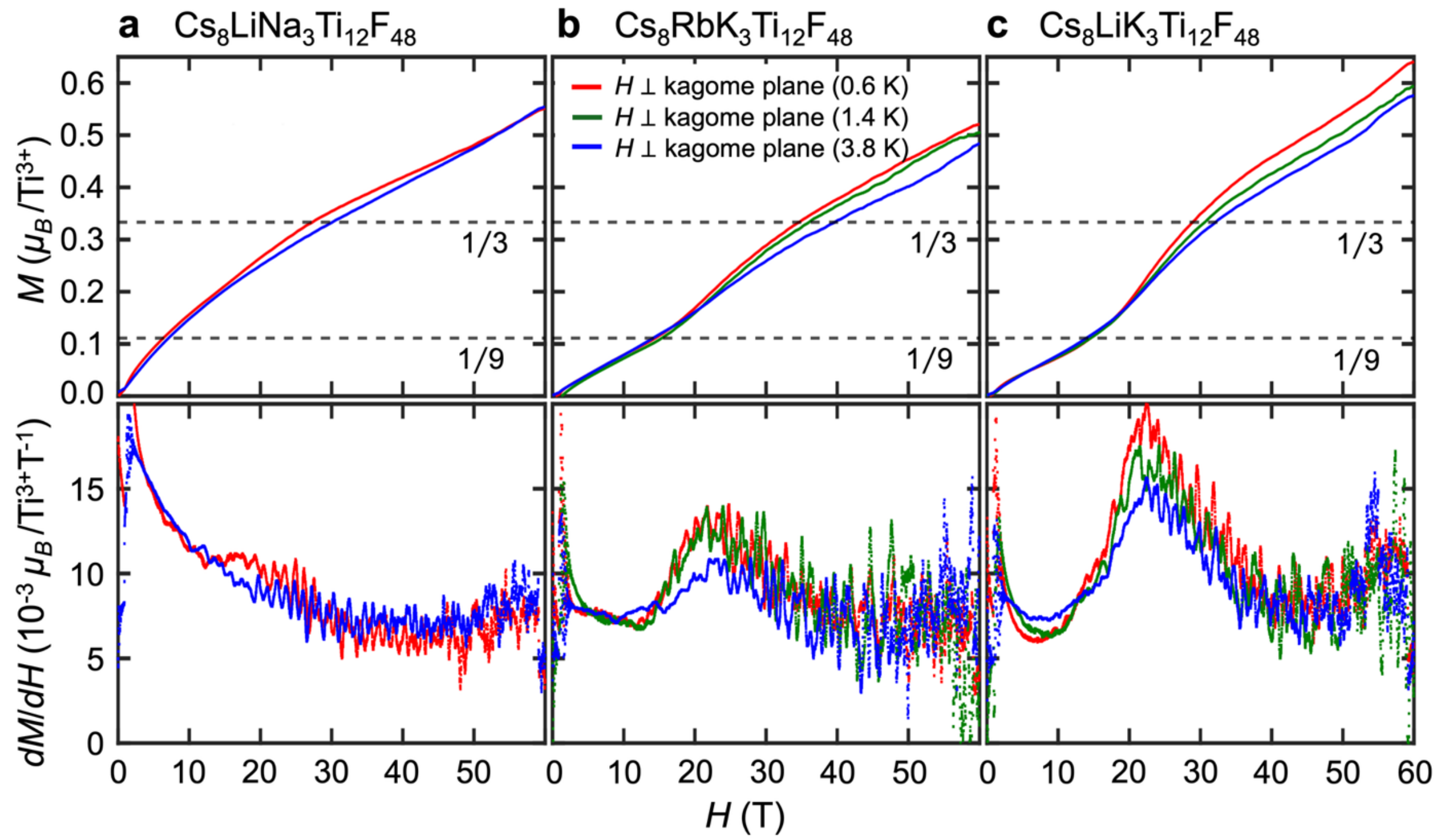

**Supplementary Fig. 4 | Temperature dependence of the high-field magnetization.** Magnetization of **a**, $Cs_8LiNa_3Ti_{12}F_{48}$, **b,** $Cs_8RbK_3Ti_{12}F_{48}$, and **c**, $Cs_8LiK_3Ti_{12}F_{48}$ measured with the field $H \perp$ kagome plane at 0.6 K (red), 1.4 K (green), and 3.8 K (blue). The dashed horizontal lines indicate the nominal 1/9 and 1/3 fractional magnetization levels. The bottom panels show the corresponding field derivative, $dM/dH$. No plateau-like feature is observed for LiNa over the entire temperature range. In contrast, the characteristic 1/9 plateau-like features in RbK and LiK broaden with increasing temperature but remain discernible up to 3.8 K.

To evaluate the temperature dependence of the field-induced magnetization features observed at 0.6 K, magnetization measurements were performed up to 3.8 K with $H \perp$ kagome plane. Supplementary Fig. 3 shows the temperature dependence of $M(H)$ and $dM/dH$ for all three compounds. Consistent with the absence of a fractional magnetization plateau, the structurally compressed LiNa compound exhibits a comparatively smooth temperature dependence. Across the measured field range up to 60 T, the magnetization increases monotonically, with the 3.8 K response remaining below the 0.6 K curve owing to thermal suppression of the field-induced magnetization.

The expanded RbK and LiK compounds exhibit a distinctly different temperature evolution. At 0.6 K, the plateau-like field ranges identified in the high-field magnetization data extend from approximately 2.5 to 15.2 T for RbK and from 2.5 to 16.6 T for LiK. With increasing temperature, these intermediate-field features progressively broaden, particularly near their upper-field boundaries. Beyond the plateau-like regime, the magnetization at 0.6 K increases more rapidly than at 1.4 K and 3.8 K, indicating that the crossover out of the plateau-like regime is temperature sensitive.

Correspondingly, the minima in $dM/dH$ become deeper and more sharply defined upon cooling to 0.6 K. Although the features broaden with increasing temperature, the characteristic minima remain discernible up to 3.8 K. These results demonstrate that the 1/9 plateau-like signatures persist up to 3.8 K.

**SUPPLEMENTARY NOTE 4: TOPAZ single-crystal neutron diffraction and crystal structure refinement**

Single-crystal neutron diffraction data for $Cs_8LiNa_3Ti_{12}F_{48}$, $Cs_8RbK_3Ti_{12}F_{48}$, and $Cs_8LiK_3Ti_{12}F_{48}$ were collected at 300 K using the TOPAZ time-of-flight (TOF) diffractometer at the Spallation Neutron Source [1]. The crystals were mounted on aluminum pins and positioned on the instrument's ambient-temperature goniometer. TOPAZ employs a broad neutron wavelength band of 0.4–3.5 Å and an array of 25 area detectors covering approximately 3.2 sr, enabling efficient three-dimensional mapping of reciprocal space. Data were collected at multiple goniometer orientations to achieve high redundancy and broad reciprocal-space coverage of the Bragg reflections.

The extensive reciprocal-space coverage was particularly important for accurately determining the positions of the $F^-$ ligands surrounding the $Ti^{3+}$ ions, which form the $TiF_6$ octahedra and define the kagome-related magnetic lattice. Initial data reduction, including peak integration and Lorentz and absorption corrections, was performed using the Mantid software package. The resulting wavelength-resolved integrated intensities were subsequently used for crystallographic refinement.

Structural refinements of the TOPAZ data were performed using JANA2020 [2] against $F^2$. Here, $F$ denotes the crystallographic structure-factor amplitude, while $F^2 = |F^2|$ is the calculated squared structure-factor amplitude and is directly related to the measured Bragg intensity. The integrated TOPAZ intensities were processed in HKLF 2 format to retain the wavelength information associated with each reflection in the TOF Laue dataset during refinement.

The structural models were refined in the polar monoclinic space group *Cm*. Because the origin is not uniquely defined in this polar space group, one reference atomic position was constrained to prevent origin drift during least-squares refinement. The Li site in the Li-containing compounds and the Rb site in the RbK compound—were therefore fixed at (0, 0, 0).

For the Li-containing compounds, the displacement parameter of Li was weakly constrained by the diffraction data and strongly correlated with other refinement parameters. The Li isotropic displacement parameter, $U_{\mathrm{iso}}$, was therefore fixed at 0.020 Å$^2$ to obtain stable refinement convergence. All remaining atomic positions and displacement parameters were refined freely, with anisotropic displacement parameters used where supported by the data.

The refined atomic coordinates, isotropic displacement parameters ($U_{\mathrm{iso}}$), and site occupancies for the three compounds are summarized in Supplementary Tables 2-4. The alphabetical suffixes a, b, and c identify crystallographically distinct sites generated by the monoclinic distortion from the corresponding parent structural sites. For a two-site splitting, the a site lies on the mirror plane and the b site occupies a general position. For a three-site splitting, the three crystallographically independent sites are labeled a, b, and c. Unsplit sites on the mirror plane are labeled a. Standard uncertainties are given in parentheses for the last significant digits.

**Supplementary Table 2 | Atomic coordinates, Wyckoff site, isotropic displacement parameters ($U_{iso}$), and site occupancies of $Cs_8LiNa_3Ti_{12}F_{48}$ at 300 K.** The structural parameters were obtained from refinement of the TOPAZ single-crystal neutron diffraction data. The numbers in parentheses are standard deviations in the last significant digits.

| Atom | Wyckoff site | $x$ | $y$ | $z$ | $U_{iso}$ (Å$^2$) | Occupancy |
|---|---|---|---|---|---|---|
| Cs1a | 2a | 0.1506(3) | 0.0000 | 0.6085(4) | 0.0266(13) | 1 |
| Cs1b | 4b | 0.39190(14) | 0.2499(4) | 0.1251(2) | 0.0269(8) | 1 |
| Cs2a | 2a | 0.6490(3) | 0.0000 | 0.6060(4) | 0.0245(13) | 1 |
| Cs3a | 2a | 0.8967(3) | 0.0000 | 0.3601(4) | 0.0333(17) | 1 |
| Cs3b | 4b | 0.63823(14) | 0.7502(3) | 0.8792(2) | 0.0227(6) | 1 |
| Cs4a | 2a | 0.3971(3) | 0.0000 | 0.3597(4) | 0.0297(15) | 1 |
| Li1a | 2a | 0.0000 | 0.0000 | 0.0000 | 0.020 | 1 |
| Na1a | 2a | 0.5003(3) | 0.0000 | 0.9948(5) | 0.0096(11) | 1 |
| Na1b | 4b | 0.28627(15) | 0.2503(3) | 0.4857(2) | 0.0094(6) | 1 |
| Ti1a | 2a | 0.2469(3) | 0.0000 | 0.9684(5) | 0.0091(14) | 1 |
| Ti1b | 4b | 0.1435(3) | 0.8733(4) | 0.2411(4) | 0.0097(9) | 1 |
| Ti2a | 4b | 0.53958(18) | 0.2493(4) | 0.5165(2) | 0.0095(7) | 1 |
| Ti2b | 4b | 0.6430(3) | 0.8743(4) | 0.2409(4) | 0.0084(8) | 1 |
| Ti2c | 4b | 0.3931(3) | 0.8758(3) | 0.7407(4) | 0.0083(8) | 1 |
| Ti3a | 2a | 0.7447(3) | 0.0000 | 0.9672(5) | 0.0093(15) | 1 |
| Ti3b | 4b | 0.3933(3) | 0.6231(4) | 0.7394(4) | 0.0100(9) | 1 |
| F1a | 4b | 0.70689(19) | 0.8993(2) | 0.0795(3) | 0.0234(8) | 1 |
| F1b | 4b | 0.35095(13) | 0.7503(3) | 0.7482(2) | 0.0248(7) | 1 |
| F1c | 4b | 0.50249(15) | 0.3427(2) | 0.6413(2) | 0.0178(8) | 1 |
| F2a | 2a | 0.1576(3) | 0.0000 | 0.3007(4) | 0.0181(10) | 1 |
| F2b | 4b | 0.20483(19) | 0.0989(2) | 0.0778(3) | 0.0271(9) | 1 |
| F3a | 4b | 0.58285(19) | 0.3496(2) | 0.4057(3) | 0.0251(9) | 1 |
| F3b | 4b | 0.62637(14) | 0.7497(3) | 0.17911(17) | 0.0185(5) | 1 |
| F3c | 4b | 0.28405(16) | 0.9080(2) | 0.8430(3) | 0.0211(8) | 1 |
| F4a | 2a | 0.9352(2) | 0.0000 | 0.7359(4) | 0.0235(12) | 1 |
| F4b | 4b | 0.78370(17) | 0.9067(2) | 0.8426(3) | 0.0214(8) | 1 |
| F5a | 2a | 0.8563(3) | 0.0000 | 0.0573(4) | 0.0269(16) | 1 |
| F5b | 4b | 0.46180(18) | 0.6396(2) | 0.8940(2) | 0.0212(8) | 1 |
| F6a | 2a | 0.4352(3) | 0.0000 | 0.7331(5) | 0.0275(14) | 1 |

| F6b | 4b | 0.5808(2) | 0.1520(2) | 0.4024(2) | 0.0251(9) | 1 |
|---|---|---|---|---|---|---|
| F7a | 4b | 0.65479(12) | 0.2499(3) | 0.59160(19) | 0.0284(7) | 1 |
| F7b | 4b | 0.7540(2) | 0.8483(2) | 0.3189(3) | 0.0272(10) | 1 |
| F7c | 4b | 0.46183(18) | 0.8605(3) | 0.8911(2) | 0.0217(9) | 1 |
| F8a | 2a | 0.3546(3) | 0.0000 | 0.0574(4) | 0.0277(15) | 1 |
| F8b | 4b | 0.25519(16) | 0.8480(2) | 0.3152(3) | 0.0214(8) | 1 |
| F9a | 2a | 0.6558(3) | 0.0000 | 0.3005(3) | 0.0203(11) | 1 |
| F9b | 4b | 0.50154(16) | 0.84331(19) | 0.6381(2) | 0.0195(8) | 1 |
| F10a | 4b | 0.52890(18) | 0.0985(2) | 0.1633(3) | 0.0219(8) | 1 |
| F10b | 4b | 0.42735(13) | 0.7480(3) | 0.42891(19) | 0.0269(7) | 1 |
| F10c | 4b | 0.32393(18) | 0.1118(2) | 0.5894(2) | 0.0249(9) | 1 |
| F11a | 2a | 0.6297(3) | 0.0000 | 0.8873(4) | 0.0289(16) | 1 |
| F11b | 4b | 0.82420(17) | 0.1119(2) | 0.5900(2) | 0.0251(9) | 1 |
| F12a | 2a | 0.1275(3) | 0.0000 | 0.8872(4) | 0.0286(14) | 1 |
| F12b | 4b | 0.52890(18) | 0.5968(2) | 0.1643(3) | 0.0236(9) | 1 |

**Supplementary Table 3 | Atomic coordinates, Wyckoff site, isotropic displacement parameters ($U_{iso}$), and site occupancies of $Cs_8RbK_3Ti_{12}F_{48}$ at 300 K.** The structural parameters were obtained from refinement of the TOPAZ single-crystal neutron diffraction data. The numbers in parentheses are standard deviations in the last significant digits.

| Atom | Wyckoff site | $x$ | $y$ | $z$ | $U_{iso}$ (Å$^2$) | Occupancy |
|---|---|---|---|---|---|---|
| Cs1a | 2a | 0.12230(12) | 0.0000 | 0.6082(3) | 0.0229(7) | 1 |
| Cs1b | 4b | 0.37468(13) | 0.74041(13) | 0.1205(3) | 0.0364(6) | 1 |
| Cs2a | 2a | 0.63771(18) | 0.0000 | 0.6067(4) | 0.0367(10) | 1 |
| Cs3a | 2a | 0.88734(12) | 0.0000 | 0.3724(3) | 0.0363(9) | 1 |
| Cs3b | 4b | 0.64338(13) | 0.75150(13) | 0.8713(2) | 0.0315(6) | 1 |
| Cs4a | 2a | 0.39534(19) | 0.0000 | 0.3734(4) | 0.0336(9) | 1 |
| Rb1a | 2a | 0.0000 | 0.0000 | 0.0000 | 0.0727(11) | 1 |
| Ti1a | 2a | 0.2528(2) | 0.0000 | 0.9922(3) | 0.0123(7) | 1 |
| Ti1b | 4b | 0.14542(17) | 0.12623(15) | 0.2537(3) | 0.0143(6) | 1 |
| Ti2a | 4b | 0.51227(15) | 0.74876(14) | 0.4992(3) | 0.0087(4) | 1 |
| Ti2b | 4b | 0.62171(14) | 0.12661(13) | 0.2311(3) | 0.0180(6) | 1 |
| Ti2c | 4b | 0.37354(15) | 0.12353(13) | 0.7313(3) | 0.0162(6) | 1 |
| Ti3a | 2a | 0.7541(2) | 0.0000 | 0.9925(4) | 0.0153(7) | 1 |
| Ti3b | 4b | 0.39428(16) | 0.37634(15) | 0.7560(3) | 0.0139(6) | 1 |

| K1a | 2a | 0.49852(15) | 0.0000 | 0.9786(3) | 0.0113(6) | 1 |
|---|---|---|---|---|---|---|
| K1b | 4b | 0.26496(12) | 0.74304(8) | 0.5008(2) | 0.0107(4) | 1 |
| F1a | 4b | 0.34346(17) | 0.25214(11) | 0.7659(3) | 0.0520(9) | 1 |
| F1b | 4b | 0.48414(10) | 0.65873(9) | 0.62872(19) | 0.0264(5) | 1 |
| F1c | 4b | 0.6943(2) | 0.90748(12) | 0.0832(3) | 0.0846(11) | 1 |
| F2a | 2a | 0.14327(19) | 0.0000 | 0.3162(3) | 0.0298(7) | 1 |
| F2b | 4b | 0.18441(17) | 0.90964(13) | 0.0844(2) | 0.0405(6) | 1 |
| F3a | 4b | 0.59831(9) | 0.66255(9) | 0.42323(17) | 0.0194(4) | 1 |
| F3b | 4b | 0.6552(2) | 0.25074(11) | 0.1995(3) | 0.0667(10) | 1 |
| F3c | 4b | 0.31120(17) | 0.09528(13) | 0.8900(3) | 0.0608(8) | 1 |
| F4a | 2a | 0.93906(15) | 0.0000 | 0.7460(3) | 0.0196(6) | 1 |
| F4b | 4b | 0.8039(2) | 0.09255(17) | 0.8838(4) | 0.0782(11) | 1 |
| F5a | 2a | 0.8482(2) | 0.0000 | 0.0927(4) | 0.137(3) | 1 |
| F5b | 4b | 0.47151(15) | 0.35067(11) | 0.88589(19) | 0.0321(5) | 1 |
| F6a | 2a | 0.40952(14) | 0.0000 | 0.7056(3) | 0.0276(7) | 1 |
| F6b | 4b | 0.54406(13) | 0.84336(11) | 0.3734(2) | 0.0299(5) | 1 |
| F7a | 4b | 0.60795(10) | 0.79116(12) | 0.59939(19) | 0.0241(4) | 1 |
| F7b | 4b | 0.72031(15) | 0.11305(16) | 0.3429(4) | 0.0865(14) | 1 |
| F7c | 4b | 0.48172(15) | 0.13737(16) | 0.8179(3) | 0.0464(7) | 1 |
| F8a | 2a | 0.3440(2) | 0.0000 | 0.1032(4) | 0.084(2) | 1 |
| F8b | 4b | 0.25889(11) | 0.13467(16) | 0.3191(3) | 0.0546(9) | 1 |
| F9a | 2a | 0.58733(14) | 0.0000 | 0.2562(3) | 0.0276(6) | 1 |
| F9b | 4b | 0.43055(11) | 0.16202(10) | 0.57347(19) | 0.0221(4) | 1 |
| F10a | 4b | 0.51284(16) | 0.86874(19) | 0.1420(4) | 0.0831(13) | 1 |
| F10b | 4b | 0.42191(10) | 0.29136(11) | 0.39828(19) | 0.0253(4) | 1 |
| F10c | 4b | 0.26737(13) | 0.88250(16) | 0.6531(3) | 0.0424(6) | 1 |
| F11a | 2a | 0.6560(2) | 0.0000 | 0.8881(3) | 0.125(3) | 1 |
| F11b | 4b | 0.81251(12) | 0.89879(13) | 0.6274(3) | 0.0369(6) | 1 |
| F12a | 2a | 0.1544(2) | 0.0000 | 0.8705(3) | 0.091(2) | 1 |
| F12b | 4b | 0.52763(14) | 0.61312(17) | 0.1927(2) | 0.0458(7) | 1 |

**Supplementary Table 4 | Atomic coordinates, Wyckoff site, isotropic displacement parameters ($U_{\mathrm{iso}}$), and site occupancies of $Cs_8LiK_3Ti_{12}F_{48}$ at 300 K.** The structural parameters were obtained from refinement of the TOPAZ single-crystal neutron diffraction data. The numbers in parentheses are standard deviations in the last significant digits.

| Atom | Wyckoff site | $x$ | $y$ | $z$ | $U_{\mathrm{iso}}$ (Å$^2$) | Occupancy |
|---|---|---|---|---|---|---|
| Cs1a | 2a | 0.1114(7) | 0.0000 | 0.6194(12) | 0.034(4) | 1 |
| Cs1b | 4b | 0.8596(3) | 0.2491(7) | 0.1032(4) | 0.0294(13) | 1 |
| Cs2a | 2a | 0.6105(7) | 0.0000 | 0.6178(12) | 0.029(3) | 1 |
| Cs3a | 2a | 0.8700(7) | 0.0000 | 0.3876(9) | 0.027(3) | 1 |
| Cs3b | 4b | 0.6178(3) | 0.2509(8) | 0.8678(5) | 0.0352(16) | 1 |
| Cs4a | 2a | 0.3671(6) | 0.0000 | 0.3846(11) | 0.030(3) | 1 |
| Ti1a | 2a | 0.2632(6) | 0.0000 | 1.0115(11) | 0.013(3) | 1 |
| Ti1b | 4b | 0.1166(6) | 0.1239(7) | 0.2429(10) | 0.015(2) | 1 |
| Ti2a | 4b | 0.9745(3) | 0.2493(7) | 0.4753(5) | 0.0124(12) | 1 |
| Ti2b | 4b | 0.6182(6) | 0.1243(7) | 0.2434(9) | 0.0116(18) | 1 |
| Ti2c | 4b | 0.3680(6) | 0.1249(7) | 0.7440(9) | 0.0113(19) | 1 |
| Ti3a | 2a | 0.7599(6) | 0.0000 | 1.0098(10) | 0.012(3) | 1 |
| Ti3b | 4b | 0.3661(6) | 0.3753(7) | 0.7414(10) | 0.015(2) | 1 |
| Li1a | 2a | 0.0000 | 0.0000 | 0.0000 | 0.020 | 1 |
| K1a | 2a | 0.5018(5) | 0.0000 | 0.9906(8) | 0.0079(19) | 1 |
| K1b | 4b | 0.2303(3) | 0.2489(6) | 0.4940(4) | 0.0094(10) | 1 |
| F1a | 2a | 0.8596(6) | 0.0000 | 0.1177(10) | 0.047(4) | 1 |
| F1b | 4b | 0.4861(4) | 0.3931(4) | 0.7938(6) | 0.0267(17) | 1 |
| F2a | 2a | 0.5759(5) | 0.0000 | 0.2470(9) | 0.022(2) | 1 |
| F2b | 4b | 0.4144(5) | 0.1569(4) | 0.5761(6) | 0.0413(18) | 1 |
| F3a | 4b | 0.9070(3) | 0.1580(4) | 0.5675(6) | 0.0264(14) | 1 |
| F3b | 4b | 0.7101(5) | 0.0970(6) | 0.1151(8) | 0.042(3) | 1 |
| F3c | 4b | 0.3707(3) | 0.2497(5) | 0.8005(3) | 0.0291(11) | 1 |
| F4a | 2a | 0.8585(5) | 0.0000 | 0.6814(8) | 0.024(2) | 1 |
| F4b | 4b | 0.8254(5) | 0.0932(5) | 0.9119(7) | 0.031(2) | 1 |
| F5a | 4b | 0.5270(5) | 0.3448(5) | 0.3707(9) | 0.043(2) | 1 |
| F5b | 4b | 0.8240(5) | 0.4057(4) | 0.9113(7) | 0.030(2) | 1 |
| F5c | 4b | 0.1614(2) | 0.2497(5) | 0.2394(4) | 0.0208(10) | 1 |
| F6a | 4b | 0.8784(3) | 0.2562(8) | 0.3661(4) | 0.057(2) | 1 |
| F6b | 4b | 0.5340(4) | 0.1484(4) | 0.1168(7) | 0.0391(19) | 1 |
| F6c | 4b | 0.2535(4) | 0.1414(5) | 0.6839(7) | 0.0358(19) | 1 |
| F7a | 4b | 0.0715(3) | 0.2443(8) | 0.5827(5) | 0.050(2) | 1 |
| F7b | 4b | 0.6992(6) | 0.1021(4) | 0.3717(7) | 0.036(2) | 1 |
| F7c | 4b | 0.4864(4) | 0.1080(5) | 0.7975(7) | 0.0292(19) | 1 |
| F8a | 2a | 0.0714(6) | 0.0000 | 0.2438(9) | 0.021(2) | 1 |
| F8b | 4b | 0.7079(4) | 0.4050(5) | 0.1141(8) | 0.037(2) | 1 |
| F9a | 2a | 0.3621(6) | 0.0000 | 0.6833(8) | 0.026(2) | 1 |
| F9b | 4b | 0.0331(5) | 0.3455(4) | 0.3773(7) | 0.045(2) | 1 |
| F10a | 2a | 0.1679(6) | 0.0000 | 0.9019(11) | 0.054(4) | 1 |
| F10b | 4b | 0.0353(4) | 0.1496(5) | 0.1090(7) | 0.0341(18) | 1 |
| F11a | 2a | 0.3626(5) | 0.0000 | 0.1194(10) | 0.043(3) | 1 |
| F11b | 4b | 0.1944(5) | 0.1020(5) | 0.3707(7) | 0.0325(17) | 1 |
| F12a | 2a | 0.6672(7) | 0.0000 | 0.8983(10) | 0.047(4) | 1 |
| F12b | 4b | 0.2502(4) | 0.3579(5) | 0.6873(7) | 0.036(2) | 1 |

**SUPPLEMENTARY NOTE 5: FONDER neutron diffraction and joint neutron and X-ray structural refinement**

To obtain an independent and more highly constrained structural models of $Cs_8RbK_3Ti_{12}F_{48}$, and additional single-crystal neutron diffraction experiment was performed at room temperature using the FONDER four-axis diffractometer at the JRR-3 research reactor of the Japan Atomic Energy Agency[3]. A monochromatic neutron beam with a wavelength of 1.2476 Å was used. Bragg reflections were measured using standard *ω*-step scans, providing high-resolution diffraction intensities for detailed structural refinement.

The FONDER neutron diffraction data were jointly refined with room-temperature single-crystat X-ray diffraction data collected using a DIP3200 X-ray diffractometer (Bruker AXS) with a Mo target. The X-ray measurements employed Mo $K\alpha$ radiation with λ = 0.71073 Å. The joint neutron and X-ray refinement was performed using JANA2020 [2], taking advantage of the complementary scattering contrasts of the two diffraction techniques. The X-ray dataset provides strong constraints on the positions and displacement parameters of the heavier Cs, Rb, K, and Ti atoms, whereas the neutron data provide complementary sensitivity to the F sublattice.

The FONDER neutron and X-ray datasets were jointly refined in the monoclinic space group *Cm* (No. 8). For the FONDER neutron dataset, the measured reflections were represented by structure-factor amplitudes *F* and their standard uncertainties, *dF,* and the refinement was performed against *F*. Dataset-specific weighting factors were applied to balance the contributions of the neutron and X-ray data in the joint least-squares refinement. Absorption and extinction corrections were applied separately to each dataset before the joint refinement. The resulting atomic coordinates define the structural model used to construct the polyhedral representation and the isolated Ti-F framework shown in main-text Figs. 1d and 1e, respectively.

**Comparison of the TOPAZ and FONDER/X-ray structural models.**

To assess the reproducibility of the structural determination, the structure obtained independently from the 300 K TOPAZ neutron diffraction data was compared with that obtained from the joint FONDER neutron/X-ray refinement.

Both refinements yield the same monoclinic *Cm* space group and the same overall atomic arrangement. The lattice parameters are closely comparable [Supplementary Table 5]. The TOPAZ refinement gives $a$ = 15.151(3) Å, $b$ = 15.238(3) Å, $c$ = 10.837(4) Å, and $\beta = 90.44(2)°$, whereas the FONDER/X-ray refinement gives $a$ = 15.169(12) Å, $b$ = 15.304(12) Å, $c$ = 10.759(10) Å, and $\beta = 90.57(8)°$ The corresponding unit-cell volumes are 2501.9(1) and 2497.5(5) Å, respectively, differing by only approximately 0.18%.

The refined Cs, Rb, K, and Ti sublattices are in good overall agreement between the two structural models, although larger site-dependent differences are observed for some individual F coordinates and displacement parameters. Thus, the independently obtained TOPAZ and FONDER/X-ray refinements provide consistent descriptions of the same *Cm* crystal structure, supporting the robustness of the symmetry assignment and the principal structural features while allowing for modest quantitative differences in some local structural parameters.

**Supplementary Table 5 | Comparison of room-temperature crystallographic parameters of $Cs_8RbK_3Ti_{12}F_{48}$ obtained from the FONDER/X-ray and TOPAZ refinements.** The FONDER-based structural parameters were obtained from the joint refinement of FONDER single-crystal neutron diffraction and single-crystal X-ray diffraction data, whereas the TOPAZ parameters were obtained independently from TOPAZ single-crystal neutron diffraction data. The relative difference is defined as $|X_{\mathrm{FONDER}} - X_{\mathrm{TOPAZ}}|/X_{\mathrm{TOPAZ}} \times 100\%$. Standard uncertainties are given in parentheses for the last significant digits.

| 300K | FONDER-based refinement | TOPAZ-based refinement | Relative difference |
|---|---|---|---|
| Space group | *Cm* | *Cm* | - |
| $a$ (Å) | 15.169(12) | 15.151(3) | 0.12% |
| $b$ (Å) | 15.304(12) | 15.238(3) | 0.43% |
| $c$ (Å) | 10.759(10) | 10.837(4) | 0.72% |
| $\alpha$ (°) | 90.0 | 90.0 | - |
| $\beta$ (°) | 90.57(8) | 90.44(2) | 0.14% |
| $\gamma$ (°) | 90.0 | 90.0 | - |
| $V$ (Å$^3$) | 2497.5(4) | 2501.9(1) | 0.17% |
| $Z$ | 2 | 2 | - |

**Supplementary Table 6 | Atomic coordinates, Wyckoff sites, isotropic displacement parameters ($U_{iso}$), and site occupancies of $Cs_8RbK_3Ti_{12}F_{48}$, obtained from joint refinement of the FONDER single-crystal neutron diffraction and X-ray diffraction data.** The same atomic-site labeling convention used in Supplementary Tables 2-4 is followed. Standard uncertainties are given in parentheses for the last significant digits.

| Atom | Wyckoff site | $x$ | $y$ | $z$ | $U_{iso}$ (Å$^2$) | Occupancy |
|---|---|---|---|---|---|---|
| Cs1a | 2a | 0.12963(6) | 0.000000 | 0.60490(8) | 0.033193(3) | 1.0 |
| Cs1b | 4b | 0.38074(4) | 0.75005(6) | 0.12271(5) | 0.034706(5) | 1.0 |
| Cs2a | 2a | 0.62956(6) | 0.000000 | 0.60514(8) | 0.032893(3) | 1.0 |
| Cs3a | 2a | 0.88605(6) | 0.000000 | 0.37133(9) | 0.034390(3) | 1.0 |
| Cs3b | 4b | 0.63794(4) | 0.75001(5) | 0.88904(5) | 0.029771(18) | 1.0 |
| Cs4a | 2a | 0.38632(6) | 0.000000 | 0.37149(9) | 0.034724(3) | 1.0 |
| Rb1a | 2a | 0.00000 | 0.000000 | 0.00000 | 0.160710(17) | 1.0 |
| Ti1a | 2a | 0.23861(12) | 0.000000 | 0.9776(2) | 0.011328(4) | 1.0 |
| Ti1b | 4b | 0.13410(13) | 0.12503(12) | 0.2471(2) | 0.012577(3) | 1.0 |
| Ti2a | 4b | 0.52707(11) | 0.75002(6) | 0.51424(14) | 0.013241(3) | 1.0 |
| Ti2b | 4b | 0.63364(13) | 0.12491(12) | 0.2465(2) | 0.011474(3) | 1.0 |
| Ti2c | 4b | 0.38316(13) | 0.12530(12) | 0.7462(2) | 0.012374(3) | 1.0 |
| Ti3a | 2a | 0.73829(12) | 0.000000 | 0.9770(2) | 0.012071(4) | 1.0 |
| Ti3b | 4b | 0.38288(13) | 0.37463(12) | 0.7468(2) | 0.011746(3) | 1.0 |
| K1a | 2a | 0.4983(2) | 0.000000 | 0.9990(4) | 0.008566(3) | 1.0 |
| K1b | 4b | 0.27181(11) | 0.75003(7) | 0.49351(17) | 0.009225(3) | 1.0 |
| F1a | 4b | 0.3364(4) | 0.2499(2) | 0.7462(5) | 0.021503(12) | 1.0 |
| F1b | 4b | 0.4694(4) | 0.6555(4) | 0.6119(5) | 0.039751(15) | 1.0 |
| F1c | 4b | 0.6759(4) | 0.9071(4) | 0.0749(5) | 0.037864(17) | 1.0 |
| F2a | 2a | 0.1411(5) | 0.000000 | 0.3066(7) | 0.025962(18) | 1.0 |
| F2b | 4b | 0.1789(4) | 0.9064(4) | 0.0780(4) | 0.033537(15) | 1.0 |
| F3a | 4b | 0.5940(4) | 0.6572(4) | 0.4190(4) | 0.025935(12) | 1.0 |
| F3b | 4b | 0.6277(5) | 0.2498(2) | 0.1880(5) | 0.029503(16) | 1.0 |
| F3c | 4b | 0.2917(4) | 0.0954(4) | 0.8731(7) | 0.041758(18) | 1.0 |
| F4a | 2a | 0.9236(4) | 0.000000 | 0.7416(7) | 0.023656(16) | 1.0 |
| F4b | 4b | 0.7881(4) | 0.0956(4) | 0.8689(7) | 0.041804(16) | 1.0 |
| F5a | 2a | 0.8316(5) | 0.000000 | 0.0845(7) | 0.049021(3) | 1.0 |
| F5b | 4b | 0.4652(4) | 0.3494(4) | 0.8713(7) | 0.038836(15) | 1.0 |
| F6a | 2a | 0.4244(4) | 0.000000 | 0.7412(7) | 0.023669(17) | 1.0 |
| F6b | 4b | 0.5935(4) | 0.8437(4) | 0.4190(4) | 0.027253(12) | 1.0 |
| F7a | 4b | 0.6232(4) | 0.7507(4) | 0.6182(5) | 0.050596(2) | 1.0 |
| F7b | 4b | 0.7502(4) | 0.1436(4) | 0.2972(5) | 0.031276(13) | 1.0 |
| F7c | 4b | 0.4666(4) | 0.1477(4) | 0.8713(7) | 0.037476(15) | 1.0 |
| F8a | 2a | 0.3339(5) | 0.000000 | 0.0853(7) | 0.046055(3) | 1.0 |
| F8b | 4b | 0.2503(4) | 0.1435(4) | 0.2962(5) | 0.031282(13) | 1.0 |
| F9a | 2a | 0.6404(5) | 0.0000 | 0.3064(7) | 0.027309(19) | 1.0 |
| F9b | 4b | 0.4705(4) | 0.1542(4) | 0.6124(5) | 0.036017(14) | 1.0 |
| F10a | 4b | 0.5154(4) | 0.8925(4) | 0.1910(5) | 0.030766(13) | 1.0 |
| F10b | 4b | 0.4318(4) | 0.2491(4) | 0.4020(7) | 0.047688(2) | 1.0 |
| F10c | 4b | 0.3047(4) | 0.8984(4) | 0.6126(5) | 0.029995(12) | 1.0 |
| F11a | 2a | 0.6387(5) | 0.000000 | 0.8631(7) | 0.043021(2) | 1.0 |
| F11b | 4b | 0.8040(4) | 0.8984(4) | 0.6132(5) | 0.031427(13) | 1.0 |
| F12a | 2a | 0.1401(5) | 0.000000 | 0.8613(7) | 0.047302(3) | 1.0 |
| F12b | 4b | 0.5155(4) | 0.6101(4) | 0.1919(5) | 0.032103(14) | 1.0 |

**SUPPLEMENTARY NOTE 6: DFT-derived magnetic exchange parameters.**

**Supplementary Table 7 | DFT-derived nearest-neighbor exchange interactions in the $Cs_8AB_3Ti_{12}F_{48}$ family.** Exchange pathways and Heisenberg couplings for $Cs_8LiNa_3Ti_{12}F_{48}$ (LiNa), $Cs_8RbK_3Ti_{12}F_{48}$ (RbK), and $Cs_8LiK_3Ti_{12}F_{48}$ (LiK), as determined from DFT-based energy mapping using the selected $U$ values shown in Fig. 4. The normalized exchange constants, $J_i/J_{max}$, are given relative to the strongest nearest-neighbor interaction in each compound. Under the Hamiltonian convention $H = \sum_{i<j} J_{ij}\mathbf{S}_i.\mathbf{S}_j$, positive and negative $J_{ij}$ denote antiferromagnetic and ferromagnetic interactions, respectively. The assignments identify exchange pathways belonging to the A and B subsystems and the corresponding zigzag connections defined in main-text Fig. 4a.

| Exchange interaction | LiNa | | RbK | | LiK | | Assignment |
|---|---|---|---|---|---|---|---|
| | $J_i$ (K) | $J_i/J_{max}$ | $J_i$ (K) | $J_i/J_{max}$ | $J_i$ (K) | $J_i/J_{max}$ | |
| $J_{2c\text{-}3b}$ | 29.6 | 0.368 | 46.8 | 0.472 | 38.4 | 0.513 | Chain-A |
| $J_{2c\text{-}2c}$ | 75.3 | 0.936 | 89.1 | 0.899 | 22.0 | 0.294 | Chain-A |
| $J_{3b\text{-}3b}$ | 74.7 | 0.928 | 99.1 | 1.000 | 32.2 | 0.430 | Chain-A |
| $J_{1b\text{-}2b}$ | 80.5 | 1.000 | 41.5 | 0.419 | 45.1 | 0.602 | Chain-B |
| $J_{1b\text{-}1b}$ | 59.7 | 0.742 | 36.2 | 0.365 | 51.7 | 0.692 | Chain-B |
| $J_{2b\text{-}2b}$ | 60.2 | 0.748 | 34.3 | 0.346 | 74.8 | 1.000 | Chain-B |
| $J_{1a\text{-}2c}$ | 40.5 | 0.503 | 41.2 | 0.415 | 42.8 | 0.572 | Chain-A-zigzag |
| $J_{2a\text{-}2c}$ | 60.2 | 0.749 | 33.0 | 0.333 | 46.8 | 0.626 | Chain-A-zigzag |
| $J_{2a\text{-}3b}$ | 52.8 | 0.656 | 51.9 | 0.524 | 45.0 | 0.602 | Chain-A-zigzag |
| $J_{3a\text{-}3b}$ | 37.7 | 0.469 | 26.9 | 0.272 | 46.2 | 0.618 | Chain-A-zigzag |
| $J_{1a\text{-}1b}$ | 7.9 | 0.098 | 53.9 | 0.544 | 29.3 | 0.392 | Chain-B-zigzag |
| $J_{1b\text{-}2a}$ | -6.3 | -0.078 | 53.2 | 0.537 | 50.9 | 0.680 | Chain-B-zigzag |
| $J_{2a\text{-}2b}$ | 9.9 | 0.123 | 38.3 | 0.387 | 40.8 | 0.546 | Chain-B-zigzag |
| $J_{2b\text{-}3a}$ | -5.1 | -0.064 | 52.2 | 0.527 | 7.87 | 0.105 | Chain-B-zigzag |

**SUPPLEMENTARY NOTE 7: Temperature dependence of high-field magnetization in powder samples.**

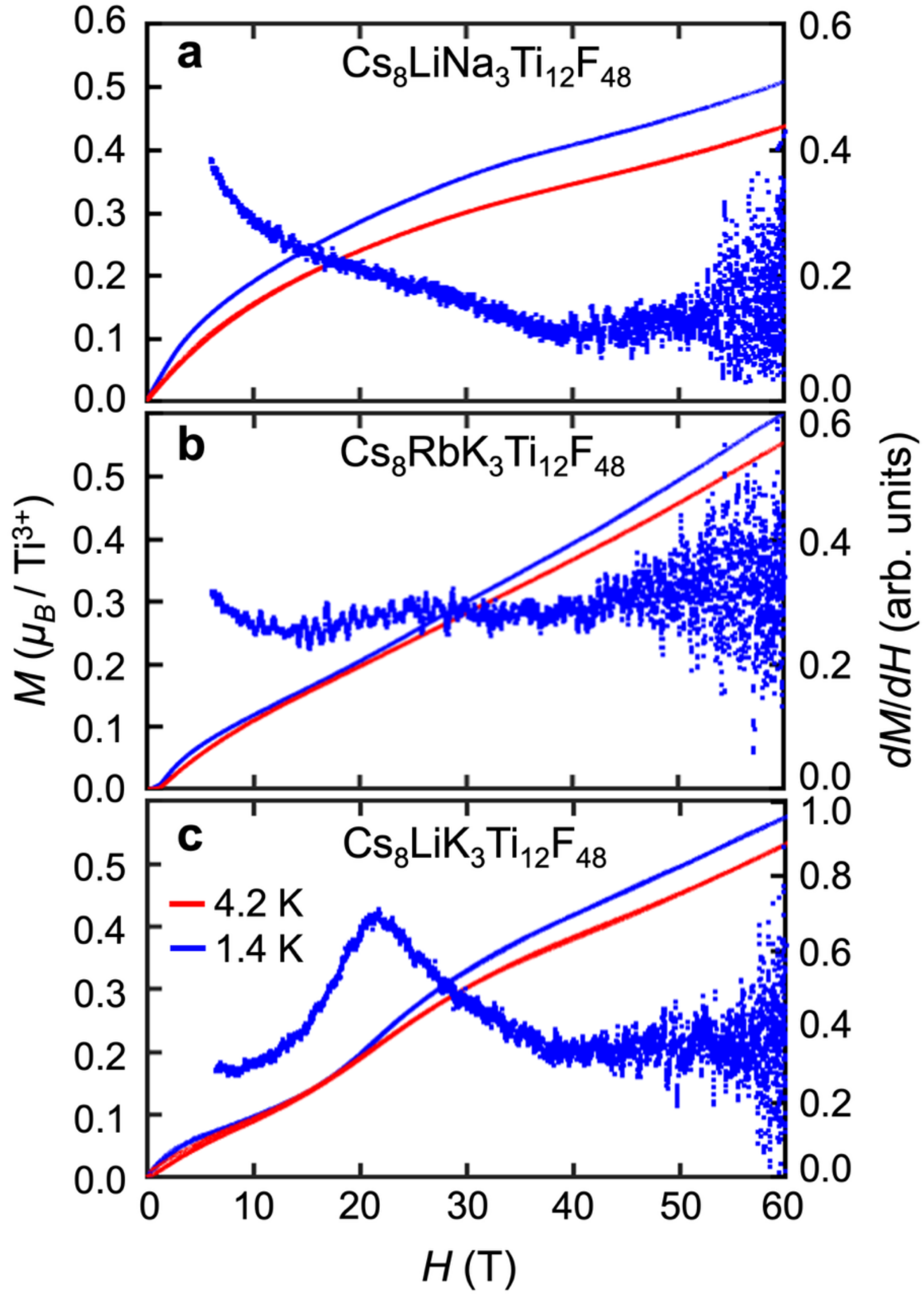

**Supplementary Fig. 5 | Temperature dependence of high-field magnetization in powder samples of the $Cs_8AB_3Ti_{12}F_{48}$ family.** Field-dependent magnetization, $M(H)$ (left y-axis), and the corresponding field derivative, $dM/dH$ (right y-axis) measured in pulsed magnetic fields up to 60 T for **a,** $Cs_8LiNa_3Ti_{12}F_{48}$ **b,** $Cs_8RbK_3Ti_{12}F_{48}$ and **c,** $Cs_8LiK_3Ti_{12}F_{48}$. Magnetization data collected at 4.2 K and 1.4 K are shown by the red and blue solid lines, respectively, while the blue points show $dM/dH$ at 1.4 K. The RbK and LiK compounds exhibit pronounced field-dependent features in $dM/dH$ within the intermediate-field regime, whereas the LiNa compound shows a comparatively smooth field dependence. Measurements were performed on powder samples at the International MegaGauss Science Laboratory, Institute for Solid State Physics, University of Tokyo.